\documentclass[12pt]{article}
\pdfoutput=1
\usepackage{arxiv}
\usepackage{amssymb,amsthm,amsmath,mathrsfs,algorithm, algorithmic}

\usepackage{latexsym,graphicx,url,enumerate,fancyhdr}
\usepackage[bookmarks=false]{hyperref}
\usepackage{url}            
\usepackage{booktabs}       
\usepackage{amsfonts}       
\usepackage{nicefrac}       
\usepackage{microtype}      
\usepackage{lipsum}		
\usepackage{parskip}
\usepackage[utf8]{inputenc} 
\usepackage[T1]{fontenc}

\newtheorem{example}{Example} 
\newtheorem{theorem}{Theorem}
 
\newtheorem{proposition}[theorem]{Proposition} 
\newtheorem{remark}[theorem]{Remark}

\newtheorem{definition}[theorem]{Definition}

\usepackage{xr}
\usepackage{comment}

\newcommand{\be}{\begin{equation}}
\newcommand{\ee}{\end{equation}}
\newcommand{\beqa}{\begin{eqnarray*}}
\newcommand{\eeqa}{\end{eqnarray*}}
\newcommand{\beqn}{\begin{eqnarray}}
\newcommand{\eeqn}{\end{eqnarray}}
\newcommand{\ba}{\begin{array}}
\newcommand{\ea}{\end{array}}
\newcommand{\bc}{\begin{center}}
\newcommand{\ec}{\end{center}}
\newcommand{\btab}{\begin{tabular}}
\newcommand{\etab}{\end{tabular}}

\newcommand{\lb}{\label}

\newcommand{\nn}{\nonumber}

\title{FLIP: A  Utility Preserving Privacy Mechanism for Time Series}

\author{ { Tucker McElroy} \\
	Research and Methodology Directorate, U.S. Census Bureau\\
	4600 Silver Hill Road,Washington, D.C. 20233-9100, USA\\
	\texttt{tucker.s.mcelroy@census.gov} \\
	\And
	{ Anindya Roy} \\
	U.S. Census Bureau\\
	University of Maryland Baltimore County\\
    1000 Hilltop Cir, Baltimore, MD 21250\\
	\texttt{anindya@umbc.edu} 
	\And
	{ Gaurab Hore}\thanks{Alternate email: gaurab.hore25@gmail.com}  \\
	University of Maryland Baltimore County\\
    1000 Hilltop Cir, Baltimore, MD 21250\\
	\texttt{gaurabh1@umbc.edu}
	}

\renewcommand{\headeright}{}
\renewcommand{\undertitle}{}
\date{}
\begin{document}
\maketitle

\begin{abstract}
Guaranteeing privacy in released data is an important goal for data-producing agencies. There has been extensive research on developing suitable privacy mechanisms in recent years. Particularly notable is the idea of noise addition with the guarantee of differential privacy. There are, however, concerns about compromising data utility when very stringent privacy mechanisms are applied. Such compromises can be quite stark in correlated data, such as time series data.  Adding white noise to a stochastic process may significantly change the correlation structure, a facet of the process that is essential to optimal prediction. We propose 
the use of all-pass filtering as a 
privacy mechanism for regularly sampled time series data, showing that this procedure preserves utility while also providing sufficient privacy guarantees to entity-level time series.
\end{abstract}


\section{Introduction}
\label{sec1}

Privacy protection in data disclosure has a long history, first being formalized by the U.S. Privacy Act of 1974. Since that time, there have been numerous developments in the control of disclosure risk while processing data.  While there are many different formulations of randomized privacy mechanisms, the majority of the algorithms have been developed for independent data.  However, given the surging demand for granular published time series in different economic sectors, statistical agencies face the need for proper disclosure avoidance algorithms that are tailored specifically to time series published over a fixed frequency, such as monthly or quarterly.  

In
\cite{Abowd2012}, the authors  propose a disclosure avoidance mechanism that relies on the use of noise infusion through a permanent multiplicative noise distortion factor, used for magnitudes, counts, differences, and ratios. That paper is 
one of the few works on privacy that investigate the perturbation of time series properties that result from application of the privacy mechanism. One of the concerns with the conventional implementation of privacy mechanisms for dependent data, such as a time series, is that the dependence structure may be significantly altered. In most applications with dependent data, the ability to estimate the dependence structure is a key facet of data utilization; for example, optimal forecasting of a time series depends upon knowing the autocorrelation structure. Thus, current privacy mechanisms result in potentially degraded data utility for time series data.

 {\it Differential privacy} (DP) is probably the most popular privacy mechanism, developed in a series of papers; see \cite{Dw2006}, \cite{DwMcNiSm2006},  \cite{DwRo2014}, for example. The concept of differential privacy ensures that the removal or addition of a single database item does not (substantially) affect the outcome of any analysis, thereby letting individual sensitive records avoid identification in a given database.
Differential privacy is algorithmically simple, and allows anything learnable via statistical queries to be learned "differentially privately". The DP framework greatly expands the applicability of output perturbation, a technique for protecting individuals' privacy by adding a small amount of random noise to the released data. DP may not be appropriate if multiple examples correspond to the same individual; a modified framework that resolves some of the hurdles in traditional privacy mechanisms is the Pufferfish framework \cite{KiMa2011}.

There is a body of work that uses modifications of DP algorithms for dependent and other structured data. Use of the Pufferfish framework for correlated data is suggested in \cite{SoWaCh2017}.  The authors introduce a generalization of DP, using the Wasserstein mechanism, which applies to any general Pufferfish framework.  In  \cite{SoCh2017}, the authors propose a modified  Pufferfish mechanism, called the Markov Quilt mechanism, and illustrated its advantages compared to conventional DP when applied to time series data. Some papers use cryptographic techniques to infuse privacy in given databases; see  \cite{Shi2011},  \cite{Hong2013}.

There is also a large literature on privacy-preserving data mining in time series; see the survey in \cite{Hong2013}.
Several perturbation methods are noise addition, compression-based perturbation, and geometric transformation perturbation. Most of the materials available on this topic discuss the aggregation method. 
In \cite{Lu2012} authors recommend an efficient privacy-preserving aggregation method based on the homomorphic Paillier cryptosystem technique, which uses a super-increasing sequence to structure multi-dimensional data. The Paillier cryptosystem can achieve homomorphic properties, widely desirable in many privacy-preserving applications \cite{SST2009}. 

Researchers are trying to reduce the risk of privacy leakage incurred by recurring components of an aggregate query, such as sums and counts of the distributed time series; \cite{SCR2011} suggests that on receiving the
query $Q$, a user evaluates $Q$ on his own time series $I_u$, perturbs the result, and sends the perturbed results
back to the aggregator. The aggregator combines the perturbed results from all users to produce the final
result. They allow both users and the aggregator to be malicious, with the flexibility that at least a fraction of users are honest. 
A framework called PASTE is proposed in \cite{RN2010} which combines the Fourier perturbation algorithm  and distributed Laplace perturbation algorithm. 
In \cite{EFM2015}, the authors consider the online setting where a user would like to continuously release  time series  data that is correlated with their own private data. 
They formalize the framework, propose a sequential scheme to achieve privacy, and finally tested the validation of the method on synthetic and real time series data. Their framework minimizes the average distance between the actual and distorted series, while
bounding the leakage of the data.

One of the earlier statistical papers that examine the privacy-utility optimization framework is \cite{WaZh2009};
the authors forcefully argue for maintaining data utility while implementing disclosure avoidance algorithms. 
In \cite{St2017}, the author describes a disclosure control model based largely on Bayesian decision theory, and defines utility and risk in the Bayesian framework. Fixing an upper bound of the risk, \cite{St2017} tries to maximize the utility.  


This body of work does not, in our opinion,
properly address the particular privacy-utility
concerns for time series data -- since both 
privacy and utility must take properties of the
temporal dynamics (such as serial correlation)
into account.  Therefore we propose a new
privacy framework (called Linear Incremental
Privacy) -- as well as new utility conditions -- for
regularly sample time series.  Our second
contribution is a novel mechanism for balancing
the horns of the privacy-utility dilemma via
the device of all-pass filtering.  This new
framework is implemented, and assessed through
both simulations and labor force data.

\section{A Privacy-Utility Framework for Time Series}
\label{sec2}

We propose a framework for making perturbations of a sensitive time series, before its release, such that the perturbations strike a balance between the conflicting objectives of retaining data utility and providing privacy protection. The main methodology involves  random phase-changed versions of the observed time series,  which leave the second-order properties of the series unchanged while providing measurable privacy protection. 

We will assume that the attacker has some prior information about the series we are trying to protect. Let $\{{\tilde{X}}_t\}$ be the series that is sensitive, and which requires protection, and let $\{{\tilde{Z}}_t\}$ be various auxiliary time series comprising the knowledge that an advanced attack can use to predict the observed series. We will assume the following observation model:
\be
\begin{pmatrix} {\tilde{X}}_t\\{\tilde{Z}}_t\end{pmatrix} = \begin{pmatrix} \mu^X_t \\ \mu^Z_t\end{pmatrix} + \begin{pmatrix} X_t \\Z_t\end{pmatrix},
\lb{eq:obs_model}
\ee
where $\{ X_t, Z_t \}$ are assumed to be jointly stationary with spectral density matrix 
\be
 f_{X,Z}(\lambda) = \begin{pmatrix}
f_X(\lambda) & f_{XZ}(\lambda)\\
f_{ZX}(\lambda) & f_Z(\lambda)
\end{pmatrix},
\lb{eq:specmat}
\ee
and $\{ \mu^X_t, \mu^Z_t \}$ are deterministic trend components that can be represented by lower order polynomials in time. We assume that the spectral matrix $f_{X,Z}$ is known to the data-publishing agency, and is also available to an {\it augury} attack.  (We use the term {\it augury} to denote a scenario that is ideal for the attacker, involving a degree of outside information.) We will first develop our proposal -- which is formulated in the frequency domain -- for a privacy-utility framework using only the stationary part $\{ X_t, Z_t \}$, and then show how to modify the proposal to incorporate deterministic trend factors  $\{ \mu^X_t, \mu^Z_t \}$.   

Before proceeding, we define some of the notations and conventions that will be used throughout. For time series, the braces notation $\{X_t\}$ will denote the entire series, while $X_t$ will denote the value of the series at time $t$. The associated Roman letter, say $X$, without the time subscript will denote the data vector. Thus, $\{Z_t\}$   denotes the underlying stationary time series representing the auxiliary information, whereas $Z = (Z_1, \ldots, Z_T)^{\prime}$ will denote the vector of the attacker's knowledge over the observation period $1, 2, \ldots, T$.  Also $z = e^{-i\lambda}$ will denote a unit norm complex variate, where $-\pi < \lambda \leq \pi$ will denote a frequency. For integrals with respect to spectral density $f$ of a function $u$, we set
$\langle u, f\rangle = {( 2 \pi )}^{-1}
\int_{-\pi}^{\pi} u(\lambda) f(\lambda) d \lambda$ and $\langle u, f\rangle_{\pi} = \pi^{-1} \int_{0}^{\pi} u(\lambda) f(\lambda) d \lambda$. Also, when $u(\lambda) = 1$, we simply denote the integrals as $\langle f\rangle$ and $\langle f \rangle_{\pi}$, respectively.  For a spectral density $f$, $\tilde{f}$ will denote the normalized density $\tilde{f}(\lambda) = f(\lambda)/\langle f \rangle_{\pi}$, and thus $\langle \tilde{f} \rangle_{\pi} =1.$

 \subsection{Post-privatization utility via all-pass filtering}

With every implementation of a disclosure avoidance mechanism, one also has to consider the utility of  privatized data -- in order to avoid nonsensical results. The utility of privatized data is achieved by maintaining analytical validity for answers to standard queries about the data.  For a time series $\{ X_t\}$, most common queries will be related to its spectral density 
$ f_X$, or the autocovariance function (ACVF) of the series, $\gamma_X(h) = \mbox{Cov}(X_t, X_{t+h}) = \langle z^h, f_X\rangle.$  Preservation of utility after privatization would therefore require that distortion of the ACVF or spectral density is minimized. The overall shift in   utility can be assessed by a discrepancy between the ACVFs---or the autocorrelation functions (ACF), denoted $\rho_X (h) = \gamma_X (h) { \gamma_X (0)}^{-1}$---of the original time series $\{ X_t \}$ and the published time series  $\{ \hat{X}_t \}$. Formally, one could measure (second order) utility by using  a normalized degradation measure 
\[U( \{ X_t \}, \{ \hat{X}_t \}) =
 1 - \mathcal{D}(f_X, f_{\hat{X}}),\]
where $\mathcal{D}$ is a discrepancy function taking values in $[0,1]$, with $\mathcal{D}(f_X, f_{\hat{X}}) = 0$ indicating no loss of utility from privatization.  Note that such a degradation measure depends only on the second order properties (i.e., the spectral density) of the original and published series, and does not depend on the sample path.  (It is possible to consider higher order utility via degradation measures associated with polyspectra, but these facets of a stochastic process are less vital than the spectral density, and will not be further pursued here.)

\begin{remark}[Noise addition and  utility] 
\label{rem:noise-infuse}
As we have seen in \cite{Joy2013}, \cite{Abowd2012}, \cite{Shi2011}, most disclosure avoidance methods for time series  have recommended noise infusion (or some related framework). While  in the usual independent scenario noise addition only affects the variance of the entity, in time series it attenuates the entire ACF, thereby significantly compromising data utility. 
Consider univariate time series for simplicity, and let $\hat{X}_t = X_t + N_t$ represent the published series, where $\{ N_t \}$ is a time series of noise infusion.  Since $\{ N_t \}$ is generated independently of the stochastic process $\{ X_t \}$, we have  
$\gamma_{\hat{X}}(h) = \gamma_X(h) + \gamma_N(h) $.  Moreover,
 since the infused noise is i.i.d., we have $\gamma_N(h)=0$ for $ |h|>0.$
 Therefore, for any lag $|h| > 0$
\[\rho_{\hat{X}} (h) = \frac{\gamma_{\hat{X}} (h)}{\gamma_{\hat{X}} (0)} = \frac{\gamma_X(h)}{\gamma_X(0) + \gamma_N(0)} < A \, \rho_X(h), \] 
where $ 0 \leq A = \mbox{SNR}/(1 + \mbox{SNR})\leq 1$, and $\mbox{SNR} = \gamma_X(0)/\gamma_N(0) > 0$ is the signal-to-noise ratio. Hence, for all $|h| > 0$ we have attenuation $ \rho_{\hat{X}} (h) < \rho_X(h).$
The amount of attenuation is directly related to   $\mbox{SNR}$, and could be substantial even for a moderately large signal-to-noise ratio.  Hence,  for time series where essential queries depend on the entire ACF, inference based on the released series $\{ \hat{X}_t \}$ could be very different from that based on the sensitive series $\{ X_t \}$, thereby severely degrading utility.
\end{remark}

It follows from Remark \ref{rem:noise-infuse} that utility is imperilled by noise infusion, no matter what the marginal structure is.  Therefore, we seek a disclosure avoidance framework that maintains utility by altering the ACVF as little as possible. We propose using convolution instead of addition; specifically, we propose a special type of linear time-invariant filtering as the primary operation for perturbing a time series.

\begin{definition}[All-pass filter]
Let $\{ X_t \}$  be a   stationary multivariate time series with spectral density matrix $f_X$. Let $B$ denote the backshift operator \cite{MP2020}, and let ${\hat{X}}_t = \Psi (B) X_t = \sum_k \psi_k X_{j-k}$ be a filtered version of $\{ X_t \}$, where $\Psi(B)= \sum_k \psi_k B^k$ is a \textit{linear time invariant filter}. Then the filter  $\Psi$ is {\it all-pass} if its frequency response function is unitary, i.e., $ \Psi (z) { \Psi (z^{-1}) }^{\prime} = I$, where $I$ is the identity matrix, and $z = e^{- i \lambda}$ for any $\lambda \in [-\pi, \pi]$.
 \end{definition} 
 
Because the spectral density matrix of $\{ \hat{X}_t \}$ is given by $ f_{\hat{X}} (\lambda) = \Psi (z)  \, f_X (\lambda) \, { \Psi (z^{-1}) }^{\prime}$, it follows that if $\Psi$ is all-pass then $\mbox{tr} f_{\hat{X}} (z) = \mbox{tr} f_X (\lambda)$.
At present we are focused on the univariate case where $|\Psi(z)| = 1$, and hence $f_{\hat{X}} (\lambda) = f_X (\lambda)$, i.e., $\mathcal{D}(f_X,f_{\hat{X}}) = 0$. Thus, perturbation of a univariate time series using an all-pass filter gives us complete   (second order) utility: $U(X,{\hat{X}}) = 1$.
For the univariate case, we can write the all-pass filter using the {\it cepstral} representation \cite{MP2020}  as
\be \Psi(z)  = \exp(\phi(z)), \quad
\phi(z)  = \sum_{k \in \mathrm{Z}} \phi_k z^k.
\lb{eq:all-pass}
\ee
The condition that $\Psi$ has to be unitary then implies that $\phi (z)$ is an anti-symmetric Laurent series, i.e., $\phi_k = - \phi_{-k}$ and $\Psi (e^{-i \lambda}) = \exp ( -i g(\lambda) )$, where $g(\lambda) = 2\sum_{k \geq 1} \phi_k  \sin (\lambda k).$ 
It follows that an all-pass filter can be thought of as a pure phase filter. In a simple all-pass filter, such as $\Psi(z) = z^a$, the filter forces a constant lag shift (by $a$ time units) for the time series. In a more general all-pass filter the phase change is frequency-dependent, and hence the effect of the filter on the time series is much more nuanced; this can be assessed through the phase delay function of the filter -- see \cite{MP2020}. 

In summary, filtering a univariate time series with a general all-pass filter provides a framework for perturbing the data that leaves  the spectral density (which encodes the second order characteristics) unaltered; whereas use of the all-pass filter preserves utility, the extent of privacy protection afforded depends upon the phase $g(\lambda)$ of the filter, as well as the stochastic properties of $\{ X_t \}$. In what follows we will refer to the all-pass filter $\Psi$ used to perturb as a {\it privacy mechanism} for $\{ X_t \}$.

\subsection{Linear Incremental Privacy (LIP)}
Noise infusion via addition is common in privacy mechanisms, because the privacy measures are typically defined in terms of the probability of change in individual data records. For time series, measures of privacy that target the marginal distribution of the random variables are not particularly useful; recall that optimal linear forecasts depend upon a process' serial correlation structure, and the marginal distribution plays no direct role (aside from its mean and variance). Put another way, successful attacks can utilize not only cross-sectional information -- which is the focus in formulations such as DP -- but also temporal information, and this latter aspect has received little attention in terms of privacy criteria.  Moreover, algorithms that optimize marginal criteria may well distort the joint dependence structure of a time series,  yielding perturbed data that no longer have the correct temporal dynamics (cf. Remark \ref{rem:noise-infuse}) -- or worse, have fallacious and spurious dynamics introduced.  

This reality motivates our proposal for a new, more suitable privacy measure for a time series; the proposal is based on an incremental notion of privacy, in terms of how prediction accuracy -- based on an adversarial information set -- is altered by a given privacy mechanism. The prediction formula is moment-based, and hence our definition is moment-based as well. 
Concisely, suppose the attacker already has some limited capacity to divine (i.e., predict) sensitive information, and we wish to measurably degrade that capability.  Since optimal prediction in statistics is classically formulated in terms of the conditional expectation (this follows from using a mean squared error loss; see \cite{MP2020}), we also adopt this paradigm in our treatment below.  We begin by considering arbitrary protection frameworks that generate a publishable $\{ \hat{X}_t \}$ from sensitive $\{ X_t \}$, and then secondly consider filtering mechanisms, and finally we specialize to the case of all-pass filters. 

For the next few paragraphs, consider a simplified scenario where
$X$, $\hat{X}$, and $Z$ are random vectors.  The best estimate --- in the sense of mean squared error (MSE) loss --- of $X$ given the attacker's information is the conditional expectation, written $E[X \vert Z]$. If $\hat{X}$ is published, then an updated attack using the additional information in $\hat{X}$ (over and above what the attacker already knew) is $E[X \vert \hat{X}, Z]$. This will be called an {\it augury} attack if the joint distribution
 of $X$, ${\hat{X}}$, and $Z$ is known. (In practice, such a joint distribution might not be known, making the attacker's task more difficult; hence the augury scenario is worst-case from the standpoint of privacy.)  In the case of linear estimators
 (which are appropriate if the random vectors are Gaussian, and
 might more generally be utilized due to their simplicity if the data is non-Gaussian),
\be
 E [ X \vert {\hat{X}}, Z ]  = E [ X \vert Z ] + \mbox{Cov} [ X, {\hat{X}} \vert Z ] \, { \mbox{Var} [ {\hat{X}} \vert Z ] }^{-1} \, ({\hat{X}} - E [ {\hat{X}} \vert Z ]).
\ee
 The second term on the right represents the update to the attack, made available by the publication of $\hat{X}$.  We say that ${\hat{X}}$ is {\it private} if this update is zero for all variables $Z$; publication of $\hat{X}$ has not assisted the attacker to predict $X$.  We distinguish between {\it augury private} and {\it feasibly private} -- the latter is a notion contingent on various possible models of the joint distributions. Computing the MSE, we obtain 
\[
  \mbox{Var} [ X \vert Z ] - \mbox{Var} [ X \vert {\hat{X}}, Z]
   =  \mbox{Cov} [ X, {\hat{X}} \vert Z ] \, { \mbox{Var} [{\hat{X}} \vert Z ] }^{-1} \,  \mbox{Cov} [{\hat{X}}, X \vert Z ].
\]
The left hand side expresses the conditional variances before and after publication of $\hat{X}$, and the difference is a measure of
additional, or incremental, vulnerability for the sensitive data.
On the right hand side of the equation we have a non-negative definite matrix -- or a non-negative scalar when $X$ is univariate.  The quantity equals zero when ${\hat{X}}$ offers no assistance to the attack.  This suggests defining
  the following quantity as the {\it privacy measure}:
\[
\mathcal{P}(X,{\hat{X}},Z)  = 1 -  \frac{ \det  \mbox{Cov} [ X, {\hat{X}} \vert Z ] \, { \mbox{Var} [{\hat{X}} \vert Z ] }^{-1} \,  \mbox{Cov} [ {\hat{X}}, X \vert Z ] }{ 
  \det  \mbox{Var} [ X \vert Z ] }.
\]
 Clearly this privacy measure is not well-defined if 
 $\det  \mbox{Var} [ X \vert Z ] = 0$, which corresponds to the
rather trivial case in which the attacker already knows the sensitive information -- in such a case
 privacy is impossible.
Otherwise, the privacy measure can be viewed as one minus a multivariate squared conditional correlation, corresponding to the $R^2$ quantity familiar from linear models.

We now specialize these notions to the case of a univariate time series; we can remove the determinant, but now the conditioning 
set is the whole time series $\{ Z_t \}$:
\begin{equation}
 \mathcal{P}( \{ X_t \},\{ \hat{X}_t \}, \{ Z_t \}) =1- \frac{\mbox{Cov} (X_t,{\hat{X}}_t \vert \{ Z_t \} )^2}{\mbox{Var}({\hat{X}}_t \vert \{ Z_t \} )
 \mbox{Var}({X}_t \vert \{ Z_t \} ) }.
\label{eq:prvcy}
\end{equation}
Although the conditional covariance and conditional variance terms in (\ref{eq:prvcy}) involve the stochastic process at time $t$,
the privacy measure is not time-dependent because of our standing assumption that $\{ X_t \}$ and $\{ Z_t \}$ are jointly stationary -- this assumption entails that these conditional quantities are the same for all $t$.

So far our privacy measure is agnostic about the  mechanism used to produce ${\hat{X}}$.  For example, we could use (\ref{eq:prvcy}) to assess noise addition; in this case, it follows from the discussion in Remark \ref{rem:noise-infuse} that $E [ \hat{X}_t \vert \{ Z_t \} ] = E [ X_t \vert \{ Z_t \} ]$ (assuming that $\{ N_t \}$ and $\{ Z_t \}$ are independent, which is natural), and hence $\{ \hat{X}_t \}$ is private.  However, due to the utility issues with noise addition, we are instead more interested in studying linear filtering mechanisms $\Psi$; the following result provides a simpler expression for the privacy measure in such a case.
  
\begin{proposition} 
Let $\{X_t, Z_t\}$ be jointly stationary with spectral matrix  \eqref{eq:specmat}. Then the error process $X_t - E [ X_t \vert \{ Z_t \} ]$ (where we have used an optimal linear predictor for the conditional expectation symbol) is stationary with mean zero and spectral density
\be
f_{X|Z}(\lambda)= f_X(\lambda)-\frac{f_{XZ}(\lambda)f_{ZX}(\lambda)}{f_Z(\lambda)}.
\lb{eq:cond_spec}
\ee
Suppose a linear filtering privacy mechanism is employed, i.e.,
 ${\hat{X}}_t = \Psi(B)X_t$ for a linear filter $\Psi$. 
 If $\langle f_{X | Z} \rangle > 0$, then
\[
\mathcal{P}( \{ X_t \},\{ \hat{X}_t \}, \{ Z_t \})
=1- \frac{\langle \Psi, f_{X|Z}\rangle^2}{ \langle  \Psi \overline{\Psi}, f_{X | Z} \rangle \, \langle f_{X|Z}\rangle },
\]
where $\overline{\Psi (e^{-i \lambda})} = \Psi (e^{i \lambda})$.
 \lb{prop:prvcy_measure}
\end{proposition}

The condition in Proposition \ref{prop:prvcy_measure} that 
 $\langle f_{X | Z} \rangle > 0$ is equivalent to saying that $X_t $ cannot be perfectly predicted from $\{ Z_t \}$; since
$\det F(\lambda) = f_Z (\lambda) \, f_{X | Z } (\lambda)$, we see that the condition fails if and only if $F(\lambda)$ is singular for $\lambda$ in a set of non-zero Lebesgue measure.  It is therefore harmless to debar this case, since it corresponds to the attacker already possessing a perfect capacity to predict the sensitive data.
 
Since the privacy measure $\mathcal{P}( \{ X_t \},\{ \hat{X}_t \}, \{ Z_t \})$  is based on protecting the residual information in $\{ X_t \}$ after its linear prediction using the attacker's information $\{ Z_t \}$, we will call the measure {\bf Linear Incremental Privacy}, and denote it by  $\mbox{LIP} (\Psi, f_{X|Z})$. More formally, we have the following definition.

\begin{definition}
Let $\{ X_t, Z_t \}$ be jointly stationary with spectral density $f_{X,Z}(\lambda)$ (\ref{eq:specmat}), and let the residual spectral density be $f_{X|Z}(\lambda)$ as defined in \eqref{eq:cond_spec}. Then the {\it Linear Incremental Privacy} (LIP) of $\{X_t\}$ given $\{Z_t\}$  with respect to the linear filtering mechanism $\Psi$ is defined as 
\[ \mbox{LIP} (\Psi, f_{X |Z})  
 =  1 -  \frac{\langle \Psi, f_{X | Z}\rangle^2}{ \langle  \Psi \overline{\Psi}, f_{X | Z} \rangle \, \langle f_{X|Z}\rangle }.
\]
\end{definition}

\begin{remark}[LIP with an all-pass filter]
\label{rem:lip-all-pass}
In the special case that $\Psi$ is an all-pass filter, LIP has a simplified form.  Using the property of an all-pass filter $\Psi$ that $\Psi \overline{\Psi} = 1$, we obtain
\begin{equation}
\label{eq:lip-all-pass}
\mbox{LIP} (\Psi, f_{X |Z})  
 =  1 -  \frac{\langle \Psi, f_{X | Z}\rangle^2}{ \langle f_{X|Z}\rangle^2 }.
\end{equation}
\end{remark}

\subsection{\texorpdfstring{$\delta$}-- LIP: a Privacy-Utility  Framework}
We describe a framework for building a privacy mechanism $\Psi$ with desirable privacy and utility properties. In the augury solution, any all-pass filter $\Psi$  would guarantee perfect utility. Hence $\Psi$ should be chosen to meet the minimum privacy requirement. If there are further considerations, then the class of all-pass filters meeting the privacy requirements can be optimized to ensure the desired properties.

Given that all augury mechanisms $\Psi$ are associated with perfect utility, an immediate approach for selecting an optimum $\Psi$ would be to maximize the privacy measure $\mbox{LIP}(\Psi, f)$:
\[ 
\Psi_{opt}  = \underset{\Psi: |\Psi(z)| = 1} 
{\arg\max} \mbox{LIP} (\Psi,f). 
\]
The optimum solution, if it exists, is usually unique and hence poses a concern in the context of data protection. If the attacker happens to know $f$, then one would be able to compute the filter coefficients for $\Psi_{opt}$ (since it uniquely depends on $f$)  and invert the computation to get the original values of the sensitive
 series $\{ X_t \}$. 
It is surprising that for any given residual spectral density $f = f_{X|Z}$ the class of all-pass filters $\Psi$ giving a perfect value of one for the privacy measure $\mbox{LIP}(\Psi, f)$  is non-empty and can be parametrized by a function class. Hence, the class of optimum privacy solutions $\Psi$ could be randomly sampled from the function class, thereby providing reasonable protection. The following result shows the existence of the solution class. 
\begin{theorem}
Let $ \{ X_t, Z_t \}$ be jointly stationary with residual spectral density $f = f_{X|Z}$, defined in \eqref{eq:cond_spec}. Let ${\hat{X}}_t = \Psi(B) X_t$ be the released series using a privacy mechanism $\Psi$. Define the class of solutions for a perfect value of the privacy measure  for a spectral density $f$ as 
\be
C_{\Psi}(f) = \{ \Psi(z) = \sum_k \psi_k z^k: |\Psi(z)| = 1, \mbox{LIP}(\Psi, f) = 1\}.
\lb{eq:opt_class}
\ee
For any residual spectral density $f$, the class $C_{\Psi}(f)$ is non-empty and contains all mechanisms of the form $\Psi(e^{-i\lambda}) = \exp \{ i\pi {\tilde{R}}(F(\lambda)) \}$ where the function $R$ belongs to the class 
\begin{equation}
 \label{classR}
   \mathcal{R} = \{R:[0, 1] \to [0, 1], R(0) = 0,  R(x) + R(1 - x) = 1, 
   \, \forall \,x \in [0,1] \},  
 \end{equation}
${\tilde{R}} (\lambda)= sgn(\lambda)R(sgn(\lambda) \lambda)$, and  $F(\lambda) = \int_0^{\lambda}
 \tilde{f} (\omega) d\omega$ with  $\tilde{f}(\lambda) = f(\lambda)/\langle f \rangle_{\pi}$ denoting the normalized spectral density.
\lb{thm:perfect_prvcy}
\end{theorem}

The optimum solutions form a large class due to the possible choices of the $R$ function. The functions in the class \eqref{classR} need not be monotone; a simple class of examples is provided by the cumulative distribution function (CDF) of any symmetric Beta random variable, with parameters $a$ and $a$ 
($\mbox{Beta}(a, a)$) for some positive real number $a$. More generally one could choose $R$ equal to a mixture of Beta CDFs, e.g.,  
\[ R(x) = \sum_{j = 1}^J \frac{\alpha_j}{2} 
 [ \mbox{Beta} (x| a_j, b_j) + \mbox{Beta} 
 (x| b_j, a_j)],
\]
where $0 < \alpha_j < 1, \sum_{j=1}^J \alpha_j = 1$ and $a_1, \ldots, a_J, b_1, \ldots, b_J$ are positive real numbers. 
To meet additional privacy utility objectives, more constraints can be put on $R$, thereby restricting the class \eqref{classR}. Later we will assume that $R$ is differentiable, and will put constraints on the Lipschitz constants and the derivatives. 

One could make a randomized choice for the function $R$  over the function class $\mathcal{R}_f$ for each conditional spectral density $f = f_{X|Z}$. However, it is not clear how much variation such a randomized choice would induce in the optimum filter. If the filters are relatively less sensitive to the choice of $R$, then
an attacker with direct knowledge of $f_{X| Z}$ would be able to approximate the filter using any optimum choice corresponding to $f$.

Following the privacy literature, we propose a privacy budget $\delta$ to enrich the class of possible mechanisms. Specifically, for a given spectral density $f$ we consider  mechanisms $\Psi$ such that $\mbox{LIP}(\Psi, f) \geq 1 - \delta$ for some $0 \leq \delta < 1$.  Thus, we have the following definition for desirable privacy mechanisms when the privacy budget is $\delta.$
\begin{definition}
\label{def:lip}
A privacy mechanism $\Psi$ is $\delta-$LIP for a given spectral density $f$  and some predetermined privacy budget $0 \leq \delta < 1$ if $\mbox{LIP}(\Psi, f) \geq 1 -  \delta.$
\end{definition} 
The following result provides a sufficient condition for a privacy mechanism $\Psi$ to be $\delta-$LIP. 
\begin{theorem}
Let $ \{ X_t, Z_t \}$ be jointly stationary with residual spectral density $f = f_{X|Z}$ defined in \eqref{eq:cond_spec}. Let $h$ be a spectral density, and suppose the privacy mechanism $\Psi_h $ belongs to the class $\mathcal{C}_{\Psi}(h)$ defined in \eqref{eq:opt_class}. Specifically, let $\Psi_h(e^{-i\lambda}) = \exp \{i\pi {\tilde{R}}(H(\lambda)) \}$ where ${\tilde{R}} (\lambda) = sgn(\lambda)R(|\lambda|)$, $R \in \mathcal{R}$ defined in \eqref{classR}, and $H (\lambda)  = \int_0^{\lambda} \tilde{h} (\omega) d\omega$ is the CDF associated with the normalized spectral density $\tilde{h} (\lambda) = h(\lambda)/\langle h \rangle_{\pi}.$
In addition, assume $R \in \mathcal{R}_L$ is Lipschitz with Lipschitz constant $L_R$, i.e., $R$ belongs to the subclass 
\be
\mathcal{R}_L = \{ R \in \mathcal{R}: |R(x) - R(y)| \leq L_R |x - y|, \, \forall x, y \in [0,1], 
L_R > 0\}. 
\lb{eq:classR_lip}
\ee
For $0 \leq \delta < 1$, let
\be
\mathcal{F}_R(f,\delta) = \left\{ h: [0, \pi] \to [0, \infty), \langle h \rangle_{\pi} = \langle f \rangle_{\pi},   \underset{0 \leq \lambda \leq \pi} \sup |h(\lambda) - f(\lambda)| \leq \frac{\sqrt{\delta}\langle f \rangle_{\pi}}{L_R \pi^{2}} \right\}.
\lb{eq:f_nbhd}
\ee
Then the privacy mechanism $\Psi_h$ is $\delta-$LIP if $h \in  \mathcal{F}_R(f,\delta).$
\lb{thm:deltaLIP}
\end{theorem}
A randomized mechanism provides greater protection. The class $\mathcal{F}_R(f_{X|Z}, \delta)$ given in Theorem~\ref{thm:deltaLIP}  is a function class, and along with the choice of the $R$ function in the definition of the optimum privacy mechanism provides a sufficiently rich  class for randomization of the privacy mechanism.  Consider a probability measure $P_R$ supported on the class $\mathcal{R}_L$; given $R \sim P_R$, let $P_{h|R}$ be a conditional probability on $\mathcal{F}_R(h, \delta)$. Then a randomized choice of the privacy mechanism would be a randomly sampled value of $\Psi(e^{-i\lambda}) = \exp \{i\pi {\tilde{R}}(H(\lambda)) \}$, where ${\tilde{R}} (\lambda) = sgn(\lambda)R(sgn(\lambda) \lambda)$ and $(R, h) \sim P_R \times P_{h|R}$.

\begin{remark}
 While the class over which the spectral density $h$ used in the construction of the privacy mechanism  can be sampled is broader than $ \mathcal{F}_R(f, \delta),$ it is important to note that regardless of which function is chosen, the privacy mechanism $\Psi_h$ is still an all-pass filter and will provide full analytical validity. 
 \end{remark}

As mentioned above, the privacy mechanism will not be useful unless one can guarantee the filter coefficients for the optimum all-pass filter are different for different choices of $R$ and $h$, particularly when $h$ is chosen in the neighborhood $\mathcal{F}_R(f, \delta)$ of the true spectral density $f$. The following result shows that even for a subclass of the possible choices of $(R, h)$,  the variation in the all-pass filter coefficients can be substantial. 

\begin{theorem}
\lb{thm:delta_LIP}
Suppose $\{ X_t, Z_t \}$ are jointly stationary with residual spectral density  $f = f_{X|Z}$ defined in \eqref{eq:cond_spec}. Assume that the released data is based on  a privacy mechanism $\Psi_h(e^{-i\lambda}) = \exp\{i\pi R(H(\lambda))\}$ where the function $R$ is as in \eqref{eq:classR_lip},  satisfying 
$\underset{0 \leq \lambda \leq \pi} \sup| \pi {\tilde{f}}(\lambda) - 1|  >  \frac{\sqrt{\delta}}{L_R\pi}$ for some $\delta>0$,
and  $h$ is  of the form 
\[ 
h (\lambda) = A \, ({\tilde{f}}(\lambda) + \Delta), 
\]
where $A = \langle f\rangle_{\pi}/[1 + \pi \Delta]$ and $H(\lambda) = \int_0^\lambda {\tilde{h}}(t) dt.$
Then $\Psi_h$ is $\delta-$LIP if  $\Delta  \in (0, B]$ for 
 \be
 B = \sqrt{\delta} \, { \left( 
 L_R \pi^2\underset{0 \leq \lambda \leq \pi} \sup| \pi {\tilde{f}}(\lambda) - 1|  - \pi \sqrt{\delta} \right) }^{-1}.
 \lb{eq:Bvalue}
 \ee
Moreover if $R$ has a derivative $r$ such that $r(x) > 0$ for all $x \in (0,1) $ and $f(\lambda) > 0$ for $0 \leq \lambda \leq \pi,$ then
\be
|\Psi_h(e^{-i\lambda}) - \Psi_f(e^{-i\lambda})| \geq \frac{\alpha\pi}{2}Q(\lambda) |F(\lambda)  - \lambda/\pi|
\lb{eq:lower_bound}
\ee
where $Q(\lambda) =  \underset{x \in L(\lambda)} \min r(x)$ with $L(\lambda)$ denoting the closed line segment joining $F(\lambda)$ and $(1 - \alpha)F(\lambda) + \alpha(\lambda/\pi)$, for any $0 < \lambda < \pi$ and $\alpha = \frac{\Delta\pi}{1 + \Delta\pi}.$
\end{theorem}

If the condition $\underset{0 \leq \lambda \leq \pi} \sup| \pi {\tilde{f}}(\lambda) - 1|  >  \frac{\sqrt{\delta}}{L_R\pi}$ is violated, that would mean that the spectral density $f$ is nearly constant, and so is any shifted version $h$. In that case, the privacy mechanism should be based on spectral densities $h$ which are in the neighborhood of $h$ but not a constant shift. 

If $\Delta \approx 0$, the lower bound on the right-hand side of \eqref{eq:lower_bound} is close to zero. Therefore, when the sampled density is close to the true density, the pointwise distance in the filter is potentially minimal. If $F(\lambda) \approx \lambda/\pi$, i.e., the true spectral density is essentially constant, then the normalized density ${\tilde{h}}$ under the constant shift model is approximately equal to  ${\tilde{f}}$. In this case also there is not much variation in the filter away from the optimum choice.  However, in general, the lower bound in \eqref{eq:lower_bound} shows that the pointwise difference between the frequency response functions of the constructed filter and that of the optimum filter under the true spectral density is bounded away from zero  over  a large frequency band, thereby providing a sufficient modification to the optimum filter.

Since $f>0$, for each $\lambda > 0$ the line segment $L(\lambda)$ is a compact sub-interval of the unit interval, and hence $Q(\lambda) >0.$  However, this bound may be extremely small, making the perturbation potentially small. The worst case scenario will be that the derivative $r$ is nearly a point mass at $x = 0.5$, in which case $R$ is nearly a constant almost everywhere. Since the user is free to choose the $R$ function, situations where $R$ is nearly a perfect sigmoidal function with a steep rise at $x = 0.5$ can be avoided. 
The assumption of a positive derivative of $R$ is not necessary, but is sufficient. In  the choices that are  recommended in this article, such as symmetric mixtures of the beta CDF, $R$ is a CDF on the unit interval with fully supported density, and hence the assumption that $r$ is positive is easily met. When $R$ is a CDF corresponding to a beta mixture of beta densities symmetric about 0.5, then the density will be monotone on $L(\lambda)$, and we can replace $Q(\lambda)$ by $r(x_{\lambda})$ where $x_{\lambda} = \min \{ F(\lambda), (1 - \alpha)F(\lambda) + \alpha(\lambda/\pi) \}.$

For computational convenience, the constant shift class $h = \tilde{f} + \Delta$ provides suitable spectral densities $h$ that could be used to construct the $\delta-$LIP privacy filter $\Psi_h$. To sample such a function conditional on $f$, one could simply sample the constant $\Delta$ from a Uniform density. Specifically, if 
\[\Delta \sim \mbox{Uniform} \left[ 0, 
B \right],
\]
then $\Psi_h$ corresponding to $h = \tilde{f} + \Delta$ is a $\delta-$LIP mechanism.

\subsection{\texorpdfstring{$\delta$}--LIP mechanism  under nonstationary trend factors}
Thus far we have developed the methodology where the sensitive series $\{ X_t \}$ and the auxiliary information series $\{ Z_t \}$ are jointly stationary. 
However, in practice the data is likely to have nonstationary features. While general stochastic trend models are popular, for a given data span the trends can often be  approximated by lower order polynomials in time. In the following, we extend the proposed methodology to the deterministic trend model \eqref{eq:obs_model}, where $\mu^X_t$ and $\mu^Z_t$ are lower-order polynomials. 

If a linear filter $\Psi$ leaves a $d$th order polynomial unchanged, then we will say $\Psi$ is a {\it $d$th order trend-invariant} filter.
If the time series is trend-stationary (i.e., when  the trend is removed,  we will be left with a stationary series), we can put constraints on $\phi(z)$ -- 
and hence $ \Psi (z)$ -- to ensure a {\it trend-invariant} privacy-utility framework (i.e., preserving lower order polynomial trends along with the autocorrelation function) using a $\delta-$LIP mechanism.  

Let $\{ X_t \}$ satisfy $(1 - B)X_t = U_t$, 
where $\{ U_t \}$ is a stationary series with spectral representation 
\[ 
U_t = \int_{-\pi}^{\pi} e^{-it\lambda}d \mathcal{Z}(\lambda),
\]
 and $\mathcal{Z}(\lambda)$ is an orthogonal increment process (see \cite{MP2020}). Setting $z = e^{-i \lambda}$, the spectral representation of $\{ X_t \}$ for $t\geq 0$ is 
\beqa
 X_t &=& X_0 + \sum_{j=1}^{t} U_j 
= X_0 + \sum_{j=1}^{t} \int_{-\pi}^{\pi} z^j d\mathcal{Z}(\lambda) \\
  &=& X_0 +\int_{-\pi}^{\pi}  \sum_{j=1}^{t} z^j d\mathcal{Z}(\lambda) 
= X_0 + \int_{-\pi}^{\pi}  z\frac{z^{t} - 1}{z - 1} \, d\mathcal{Z} (\lambda). 
 \eeqa
Suppose ${\hat{X}}_t = \Psi(B)X_t = \sum_k \psi_k X_{t-k}$. Then 
\beqn
{\hat{X}}_t  &=& \sum_k \psi_k \, 
 \left( X_0 + \int_{-\pi}^{\pi} z 
  \frac{z^{t-k} - 1}{z - 1} \, d\mathcal{Z} (\lambda)
  \right) \nn \\
&=& \sum_k \psi_k X_0 + \int_{-\pi}^{\pi} z \frac{\sum_k \psi_k z^{t-k} - \sum_k \psi_k}{1 - z} \, d\mathcal{Z} (\lambda) \nn \\
&=& \Psi(1) X_0 + \int_{-\pi}^{\pi} z\frac{\Psi(z)z^{t} - \Psi(1)}{z - 1}\, d\mathcal{Z} (\lambda).
\lb{eq:filter_trend}
\eeqn
 This last expression is derived following the discussion in 
 \cite{MW2016}. Recall that $\Psi(z) = \exp \{\phi(z) \}$. 
 If $\phi (1) = 0$ (i.e.,$\Psi(1) = 1$), from \eqref{eq:filter_trend} it follows that  only the increments
  $X_t - X_{t-1}$ get their phase altered by the all-pass filter, while the initial value (and the level) remain  unchanged. There is no phase delay at frequency zero. Hence, any linear trend is passed unaltered. 
  
  Generalizing to higher order polynomial effects of order $d$, there is no  phase delay at frequency zero provided that the filter $\Psi = \exp \{\phi \}$ has $d$ vanishing derivatives at zero:
  \be
  \frac{\partial^k \phi (z)}{\partial z^k}\Big{|}_{z = 0} = 0,\quad 0 \leq k \leq d.
  \lb{eq:vanish_der}
  \ee
Thus, to construct a $d$th order trend-invariant utility preserving $\delta-$LIP mechanism with respect to an original spectral density $f = f_{X|Z}$, we need to put additional constraints on the $R$ function and the sampled spectral density $h$ so that the filter constructed as $\Psi_h(z) = \exp \{\phi(z) \} =  \exp \{ i \pi R(H(\lambda)) \}$ satisfies the derivative condition \eqref{eq:vanish_der}. This would mean that $R$ has vanishing derivatives at $\lambda = 0$, and that $h$ is bounded with the desired number of bounded derivatives at $\lambda = 0$.

\begin{theorem}
Let a  privacy budget $\delta$ be given. Assume all the conditions of Theorem~\ref{thm:delta_LIP} hold.  Then any privacy mechanism  of the form $\Psi_h(z) = 
\exp \{i \pi R(H(\lambda)) \}$,  where the pair $(R, h)$ belong to the class
\be
  \mathcal{C}_{\delta}(R, h, f) = \{(R, h):  R \in \mathcal{R}_L, h \in  \mathcal{F}_R(f, \delta), \underset{0 \leq k \leq d} \max |R^{(k)}(0)| = 0,  \underset{0 \leq k \leq d-1} \max |h^{(k)}(0)| < \infty \}, 
\lb{eq:trend_delta_LIP}
\ee
will be a  $d$th order trend-invariant utility preserving privacy mechanism. 
Here the classes $\mathcal{R}_L$ and $\mathcal{F}_R(f, \delta)$ are defined in \eqref{eq:classR_lip} and \eqref{eq:f_nbhd}, respectively, and  the derivatives $R^{(k)}(x)$ and $h^{(k)}(\lambda)$ are assumed to be well defined in an open neighborhood of zero.
\lb{thm:trend_delta_LIP}
\end{theorem}

\section{ FLIP: A Feasible version of Linear Incremental Privacy}
\label{feasibleLIP}

There are two practical issues to resolve before the LIP framework can be implemented: the calculation of filter coefficients from a given choice of spectral density $f$ (and transform $R$), and the estimation of $f$ from the data.

\subsection{Computation of the Filter Coefficients}
 
We need  to compute the filter coefficients $\{\psi_k\}$ associated with the filter $\Psi_h(z) = \exp \{ i\pi R(H(\lambda)) \}$. 
Recall that the odd function $g(\lambda) = -\pi R(H(\lambda)) = i \phi (e^{-i \lambda})$ (defined over the interval $[-\pi, \pi]$) has a Fourier expansion in terms of the cepstral coefficient $\{\phi_k\}$, viz.,
$g(\lambda)= 2\sum_{k=1}^{\infty} \phi_k \sin(\lambda k)$.  Inverting this relation,
we have 
\be
\phi_k= \frac{1}{2 \pi} \int_{-\pi}^{\pi} g(\lambda) \sin(\lambda k) d\lambda=\frac{1}{\pi} \int_{0}^{\pi} g(\lambda) \sin(\lambda k) d\lambda, \hspace{0.3cm} k=1,2,\dots.
\lb{eq:cepstrum_coeff}
\ee
These coefficients $\phi_k$ can be computed
for any desired $k$; in practice, we can
terminate the computations when $k$ is sufficiently
large such that $\phi_k$ is negligible (say,
 machine precision), or up to an order that is computationally feasible.    

Next, the filter coefficients $\psi_k$ are obtained by matching coefficients in the expansion $\Psi(z) = \sum_k \psi_k z^k = \exp{\sum_k \phi_k z^k}$. As $\phi (z)$ is a Laurent series corresponding
 to an odd sequence, we can write
 $\phi (z) = \phi^+ (z) - \phi^{+} (z^{-1})$,
  where $\phi^{+} (z) = \sum_{k \geq 1} \phi_k z^k$ is a   power series.
 Setting 
\begin{align*}
     \psi^{+} (z) & = \sum_{k \geq 0} \psi_k^{+} z^k = \exp  \{ \phi^{+} (z) \} \nn \\
    \psi^{-} (z) & = \sum_{k \geq 0} \psi_k^{-} z^k
    =    \exp  \{- \phi^{+} (z) \},
\end{align*}
  we see that  the filter can be expressed as
  \begin{equation}
  \label{eq:wh-factor}
  \Psi (z) = \exp \{ \phi^{+} (z) \} \cdot
   \exp \{ - \phi^{+} (z^{-1}) \} 
   = \psi^{+} (z) \cdot \psi^{-} (z^{-1}),
  \end{equation}
 which is a Wiener-Hopf factorization of the Laurent series $\psi (z) = \sum_k \psi_k z^k$.
 It is clear that we only need a method
 of computing $\{ \psi_1^{+}, \psi_2^{+}, \ldots \}$
 and $\{ \psi_1^{-}, \psi_2^{-}, \ldots \}$
  from $\{ \phi_1, \phi_2, \ldots \}$, and
 this is provided by the cepstral recursions
 of \cite{MP2020}.  In particular,
 setting  $\psi^{+}_{0} = 1$ and $\psi^{-}_{0} = 1$,
 the recursions are
 \beqn
 \label{eq:filter_coeff}
 (j+1)\psi^{+}_{j+1} &=& \sum_{k=0}^{j} (k+1)\phi_{k+1}\psi^{+}_{j-k},\nn \\ 
 (j+1)\psi^{-}_{j+1} &=& -\sum_{k=0}^{j} (k+1)\phi_{k+1}\psi^{-}_{j-k}
\eeqn
 for $j = 0,1,\ldots$. 
 Finally, collecting the coefficients of $z^j$ in the product (\ref{eq:wh-factor}),
the filter coefficients $\psi_j$ can be determined.

 The cepstral coefficients $\phi_k$ are obtained by computing the integrals in \eqref{eq:cepstrum_coeff} numerically. Thus, computing a large number of $\phi_k$ may be computationally expensive. In practice, a reasonable approximation to the Fourier series of $g(\lambda)$ can be attained by truncating the series at a finite order, say $K$,  and using only finitely many $\phi_k$, for $k \leq K$.
 The error in the approximation of $\phi^{+} (z) = \sum_{k\geq 1}\phi_k z^k$ by $\phi_K^{+} (z) = \sum_{1 \leq k \leq K}\phi_k z^k$ can be bounded as a function of $K$ based on the smoothness of $\phi$. For example,
\[
\| \phi^{+} - \phi_K^{+} \|_{\infty} \leq 
(2\pi (2r - 1)K^{2r-1} )^{-1}\langle g^{(r)}\rangle 
\]
for a positive integer $r$, where $g^{(r)}$ is the $r$th derivative of $g(\lambda)$ with respect to $\lambda$. Thus, one could minimize the value of the upper bound over a reasonable range of values for $K$ and choose the smallest value that minimizes the bound over that range. In our simulation, we  found 
that $K = 20$ provided an adequate approximation to the desired cepstral form of the filter. An important point to note is that regardless of the choice of $K$, the all-pass filter $\Psi_K (z) =
\exp \{ \phi_K^{+} (z) - \phi_K^{+} (z^{-1}) \}$ would still provide full analytical validity for the released data. 
  
In practice some truncation of $\Psi (z)$ is needed
when applying the filter to a time series sample
of length $T$, because it is not possible to compute
infinitely many filter coefficients. 
If for some $M \geq 1$  a two-sided filter $\Psi (z)$
has length   $2M+1$, with $M$ future and $M$ past data
points being filtered, then the filter output
$\{ \hat{X}_t \}$ will not have values for the 
first and last $M$ data points in the sample;
 see \cite{MP2020}.  This will result in a series of length $(T - 2M)$.  In order to have a final released series that is of the same length as that of the original sensitive series, we use forecasts and backcasts to extend the series by $M$ consecutive observations
 (on both the beginning and end of the sample), and then apply the filter on the extended series to obtain a final series of length $T$. Specifically, if $(X_1, \ldots, X_T)$ is the observed series, then we create an extended series $({X}^E_{-M+1}, \ldots, {X}^E_0, X_1, \ldots, X_T, {X}^E_{T+1}, \ldots, {X}^E_{T+M})$, where ${X}^E_t $ an estimate of the 
 minimum mean squared error linear projection
 (i.e., the conditional expectation 
 $E(X_t | X_1, \ldots, X_T)$ if the process is 
 Gaussian). The forecasts can be obtained using the spectral density $f_{X}$ already computed for the privacy mechanism. 
 
\subsection{Spectral Density Estimation}

The $\delta-$LIP framework is based on knowing
the true spectral density $f$; however,
it is typically necessary to estimate the spectral density (although it's possible that the data publishing agency may have historical values from prior modeling).
 We refer to the application of the 
 $\delta-$LIP framework, when 
working with a spectral density estimate $\hat{f}$,
as a Feasible $\delta-$LIP, or $\delta-$FLIP.
That is, a privacy mechanism $\Psi$ is 
 $\delta-$FLIP for a spectral density estimate
 $\hat{f}$, and some predetermined privacy budget
 $0 \leq \delta < 1$, if 
 $\mbox{LIP}(\Psi, \hat{f}) \geq 1 -  \delta.$
 This is just Definition \ref{def:lip}, where
 $\hat{f}$ is used for the spectral density.
 Since the resulting mechanism $\Psi$ depends
 on $\hat{f}$, and not the true $f$, privacy
 becomes an empirical measure; this definition is
 sensible, because the prediction paradigm that
 lies behind our privacy framework must also
 depend on spectral density estimates, and not
 on the true unknown spectrum.  
 
 As regards utility, recall that because we are using an all-pass filter for our privacy mechanism, utility is preserved automatically {\it for any} choice of $\phi (z)$ with anti-symmetric coefficients; therefore, any
statistical error made in the estimation of 
 the spectral density matrix $f_{X,Z}$
 has no bearing on utility whatsoever. 
 

To obtain a data-based estimate of the spectral density
$f_{X,Z}$ of the stationary part, one could detrend the data by fitting a $d$th order polynomial and use the residual as approximation for $\{ X_t, Z_t \}$.  Typically for a trend estimation problem, estimation of a deterministic trend would change the time series properties of the error process $\{ X_t, Z_t \}$.  However, here the trend is removed only at this step for the estimation of the spectral density of the stationary portion -- see
(\ref{eq:obs_model}) and (\ref{eq:specmat}).
Because we are implementing a $d$th order trend-invariant filter, the obtained privacy filter $\Psi$ can be directly applied to the observations ${\tilde{X}}_t$ (or their forecast-extended version, for which we need to extend the trend at either end of the series).

From the residual time series one can estimate $f_{X,Z}$ by using a parametric model class or by nonparametric methods (see discussion in 
\cite{MP2020}), as summarized in the two options\footnote{We also studied a third option, where  a model is first selected using a model selection procedure (such as AIC) within a class of parametric models (such as a Vector Autoregression), and then the required quantities are computed based on the fitted model.  The performance was similar to that of the parametric and nonparametric options considered below, and hence we do not report the results.} below:

\begin{itemize}
\item
Option 1: Estimate $f_{X,Z}$ with a model-based estimator, such as the spectral matrix for a
Vector Autoregression of order $p$, or VAR($p$). Assuming a parametric model is correctly specified, estimate the model parameters using maximum likelihood (or a similar procedure). Once the model has been fitted to the data, plug into the expression for the model spectral matrix and obtain $\hat{f}_{X,Z}$.
\item
Option 2: Use a nonparametric estimator of $f_{X,Z}$ based upon the sample autocovariances, viz.
${\hat{\Gamma}}_k = T^{-1}\sum_{t = 1}^{T-k} W_t W_{t+k}^{\prime}$ for $k \geq 0$, where $W_t = (X_t, Z_t)^{\prime}.$  For a $d$th order trend-invariant filter, the nonparametric estimator has to be chosen to have bounded derivative at $\lambda = 0$ up to order $d$. We use a flat-top taper \cite{Politis2001} with threshold $C = 1/T$ to obtain an estimate of $f_{X,Z}$ based on the sample autocovariances. 
\end{itemize}

\subsection{Implementation of FLIP }

We propose using the following steps for an implementation of the $\delta-$FLIP mechanism.

\begin{enumerate}
\item
Estimate the spectral density matrix $f_{X,Z}$ \eqref{eq:specmat} based on the observed data.

\item
Compute the conditional spectral density $f= f_{X|Z}(\lambda)$ from $f_{X,Z}$ using  \eqref{eq:cond_spec}. 
\item
For privacy budget $\delta \geq 0$ and for desired order $d$ of the trend, choose $R \in \mathcal{R}_L$ with the desired number of derivatives at zero. Compute  $B$ in \eqref{eq:Bvalue}. Sample 
\[
\Delta \sim \mbox{Uniform} [0, B],
\]
and construct the sample spectral density 
\[ 
h (\lambda) = A (\tilde{f} (\lambda) + \Delta), 
\]
where $A = \langle f\rangle_{\pi}/[1 + \pi \Delta].$
Compute the function $g(\lambda) = -\pi R(H(\lambda))$. 
\item
Compute the coefficients $\phi_k$ (for $k \leq K$) 
in the cepstral representation of the all-pass filter \eqref{eq:all-pass} using \eqref{eq:cepstrum_coeff}. 
\item
Fix truncation value $M$, and compute the all-pass filter coefficients $\psi_j$
 for $-M \leq j \leq M$ from \eqref{eq:filter_coeff}.
\item
Use the estimated $f_{X,Z}$ to extend the series
with $M$ forecasts and $M$ backcasts,
 denoted via $\{ \tilde{X}_t^E \}$.
If a trend model has been used, employ the estimated trend values to add to the extension on either end of
the sample.
\item
Apply the filter $\{\psi_k\}_{-M}^M$ to obtain the values of the privatized series 
\[ 
{\hat{X}}_t =   \sum_{j = -M}^M \psi_j \tilde{X}^E_{t-j}, \, t = 1, \ldots, T.
\]
\end{enumerate}

\section{Numerical illustrations}
In this section we present the results of a modest simulation study, as well as some real data analysis. The simulated data are a bivariate time series, with the first component $\{ X_t \}$ being the sensitive time series of interest; the second component $\{ Z_t \}$, an auxiliary correlated time series, is assumed to be what the attacker has access to. We examine different scenarios associated with varying degrees of correlation between the sensitive series and the auxiliary series. 
The real data are from the  Quarterly Workforce Indicator (QWI) database published by U.S. Census Bureau; a detailed description of the data is given in Example~\ref{QWI data} below.\\

\begin{example}
\label{exmp:example1}
\noindent{\underline{\bf FLIP mechanism with $\delta = 0.$}}\\
{\rm{
In the first example we investigate a FLIP solution with $\delta = 0$ (i.e.,  maximum privacy), and report the operating characteristics of the FLIP mechanism. Note that in this case the randomization only occurs through the selection of an $R$ function; the spectral density $h$ used in the construction of the privacy mechanism has to be the true density $f = f_{X \vert Z}$. 
The sensitive series and the attacker's series are assumed to be jointly described by a bivariate  VAR(1). Specifically, if $\{X_t\}$ is the sensitive series and $\{Z_t\}$ represents the attacker's knowledge, then the data generating model is 
\begin{equation*} \begin{pmatrix}X_t\\Z_t\end{pmatrix} = \begin{bmatrix} \Phi_{11} & \Phi_{12} \\ 
 \Phi_{21} & \Phi_{22} \end{bmatrix} \begin{pmatrix} X_{t-1} \\ Z_{t-1} \end{pmatrix} + \begin{pmatrix} \epsilon_{t}\\\zeta_{t} \end{pmatrix}, 
\end{equation*}
where $(\epsilon_{t}, \zeta_{t})^{\prime}$ are independently and identically distributed as a bivariate normal with mean zero and covariance matrix $\Sigma$ equal to a scalar $\sigma^2$ times the two dimensional identity matrix, i.e., $(\epsilon_{t}, \zeta_{t})^{\prime} \sim N_2( {\bf 0}, \sigma^2{\bf I}).$ 

We parameterize the process in terms of the cross-correlation $\rho = Corr(X_t, Z_t)$, and also assume the $\mbox{Var} (X_t) = \mbox{Var} (Z_t) = v$. This allows us to examine the impact of the attacker's knowledge, summarized in terms of the cross-correlation $\rho$, on the proposed privacy-utility framework. 
Thus, the stationary covariance matrix of $(X_t, Z_t)$ is equal to 
\[
\Gamma_0  = v\begin{bmatrix} 1&\rho\\\rho&1\end{bmatrix},
\]
where for stationarity $\Gamma_0$ must satisfy the Riccati equations $\Gamma_0 = \Phi \Gamma_0\Phi^T + \Sigma$, and $\Phi$ is the coefficient matrix. Thus, $\Gamma_0 - \Sigma$ must be positive definite;
 in order for this to happen,   the matrix
 $\Gamma_0 - \Sigma$  must have non-negative determinant -- the condition reduces to $v\geq \sigma^2/(1 - \rho).$  Here we set $v = \sigma^2/(1 - \rho) + 1$ to make $\Gamma_0 - \Sigma$ positive definite. The coefficient matrix is then solved from the Riccati equations as $\Phi = (\Gamma_0 - \Sigma)\Gamma_0^{-1/2}$, where $\Gamma_0^{-1/2}$ is a matrix square-root of $\Gamma_0^{-1}.$
 This parameterization provides a VAR(1) with cross-correlation equal to $\rho$ and error variances equal to $\sigma^2$.
 
 For the present example we chose $\sigma^2 = 0.5$. The  cross correlation was chosen to be either
 $0.1$ or $0.7$, to allow for cases representing,
 respectively, either a low or high degree of information possessed by the attacker.
We generated $500$ Monte Carlo replicates of the bivariate series in each case, with sample size $T = 200.$  

We implemented the steps for performing FLIP on each of the generated series, and the results are presented in Figures~\ref{fig:example1} and \ref{fig:prvcy_util}. For implementing FLIP we chose the truncation order for the all-pass filter to be $K =25$ for the cepstral representation and to be $M =45$ for the filter coefficients.
The forecast extension was done using both options as described earlier in the steps of the FLIP algorithm. We only show the results for option one, since both options yielded similar results. 
Figure~\ref{fig:example1} shows a typical original sensitive series (black) with the privatized version (red) for the two different cross-correlation values (top left and bottom left panels). The figure also shows the sample autocorrelation values for the original and the privatized series for the two cases (top right and bottom right panels). 

While each privatized series retains most of the dynamic features of its original version, the pointwise differences are substantial over different intervals of the observation window, showing that predictions based on the privatized version will also have substantial differences with the original values. The sample ACF plots show that most lags in the autocorrelation plots estimated based on the sensitive and the privatized series are close, and thus statistical utility is retained to a large degree. The minor discrepancies in the autocorrelation functions arise because the filter -- due to the truncation of filter coefficients at a finite lag $M$ -- is only approximately an all-pass filter. The agreement between the ACF of the original series and the privatized series should improve with larger sample size when longer filter lengths can be allowed.  

\begin{figure}[htbp!]
\centering

\includegraphics[width = 1.8in, height = 1.2in]{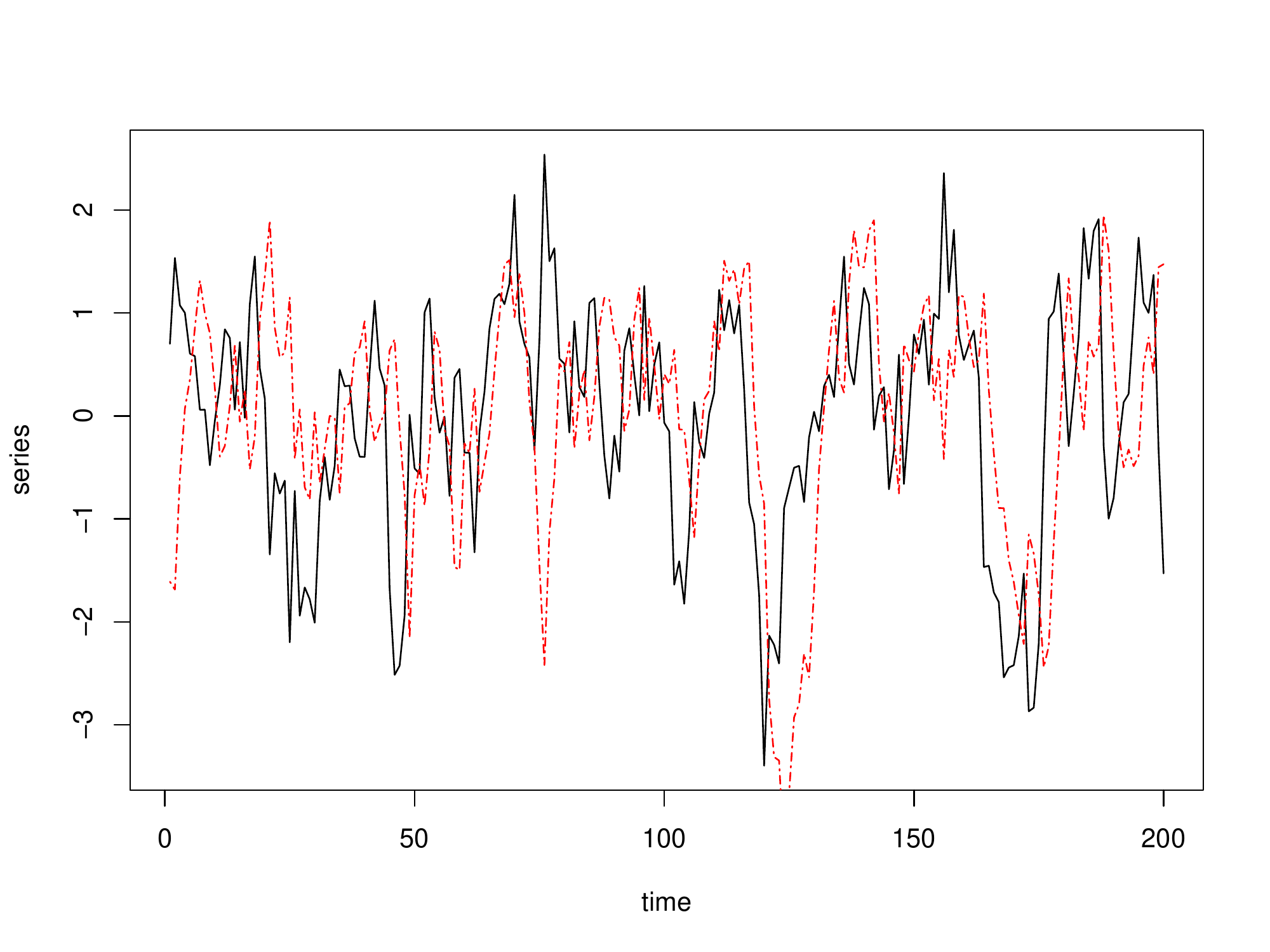}
\includegraphics[width = 1.8in, height = 1.2in]{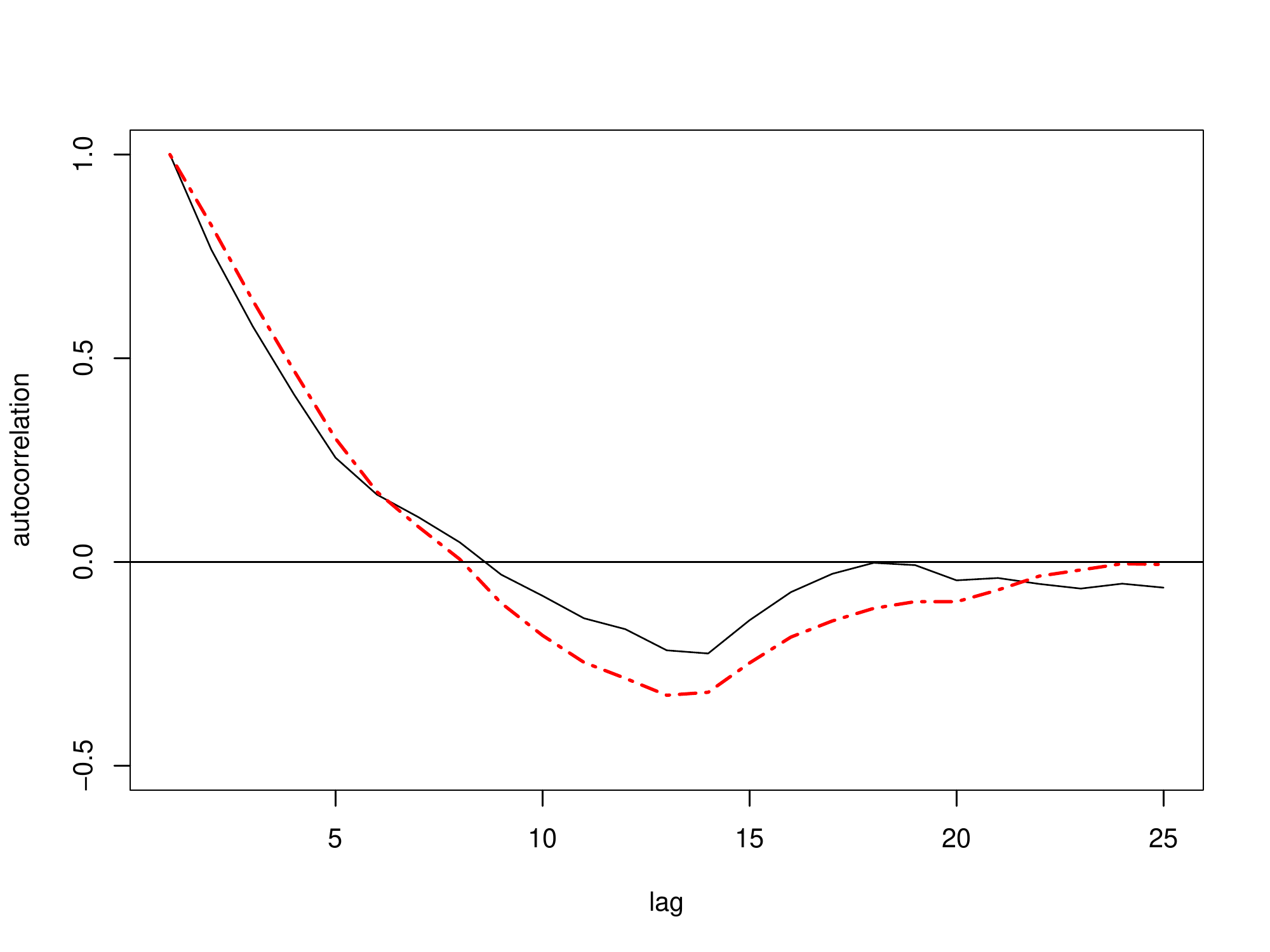}\\
\includegraphics[width = 1.8in, height = 1.2in]{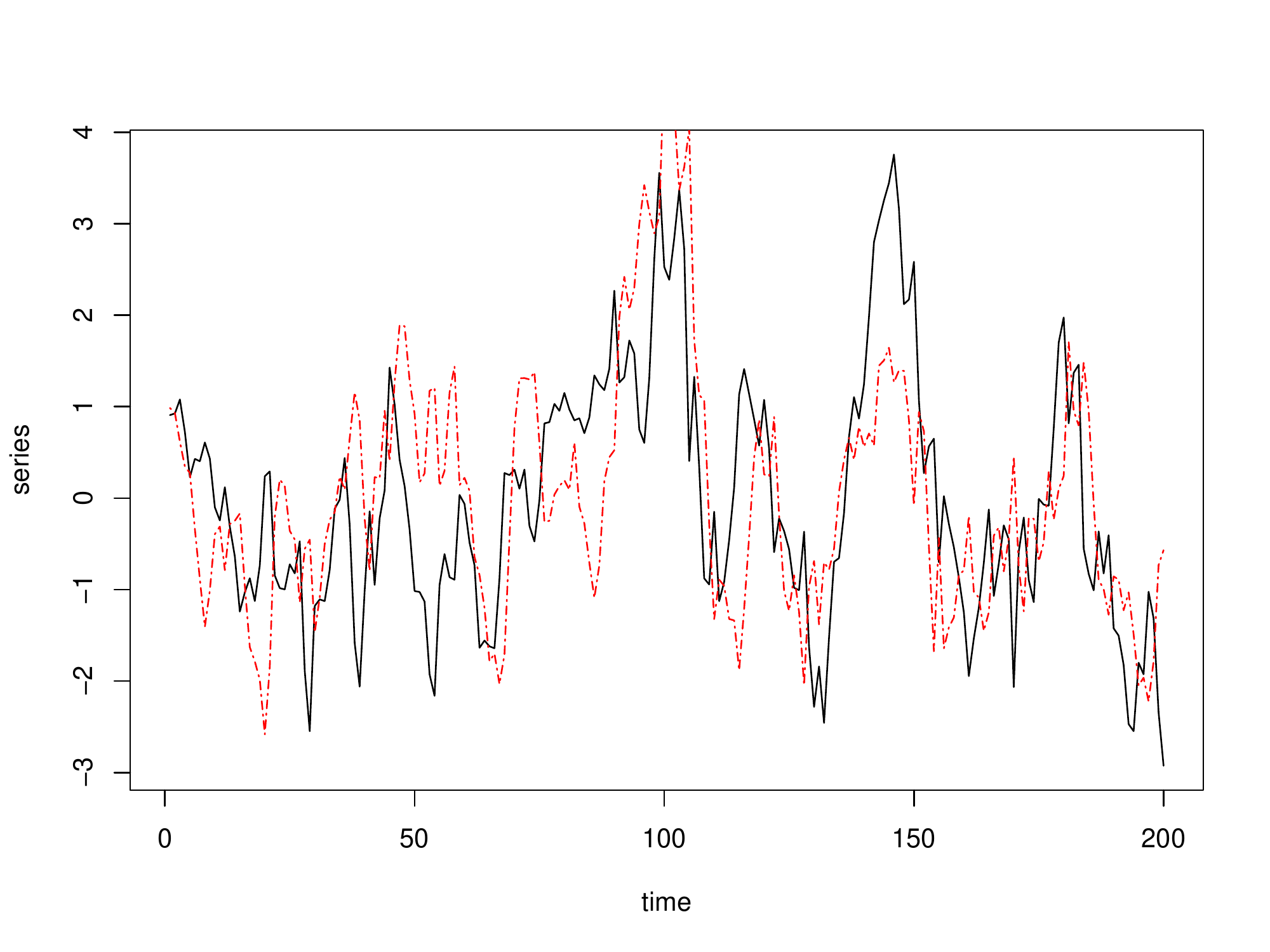}
\includegraphics[width = 1.8in, height = 1.2in]{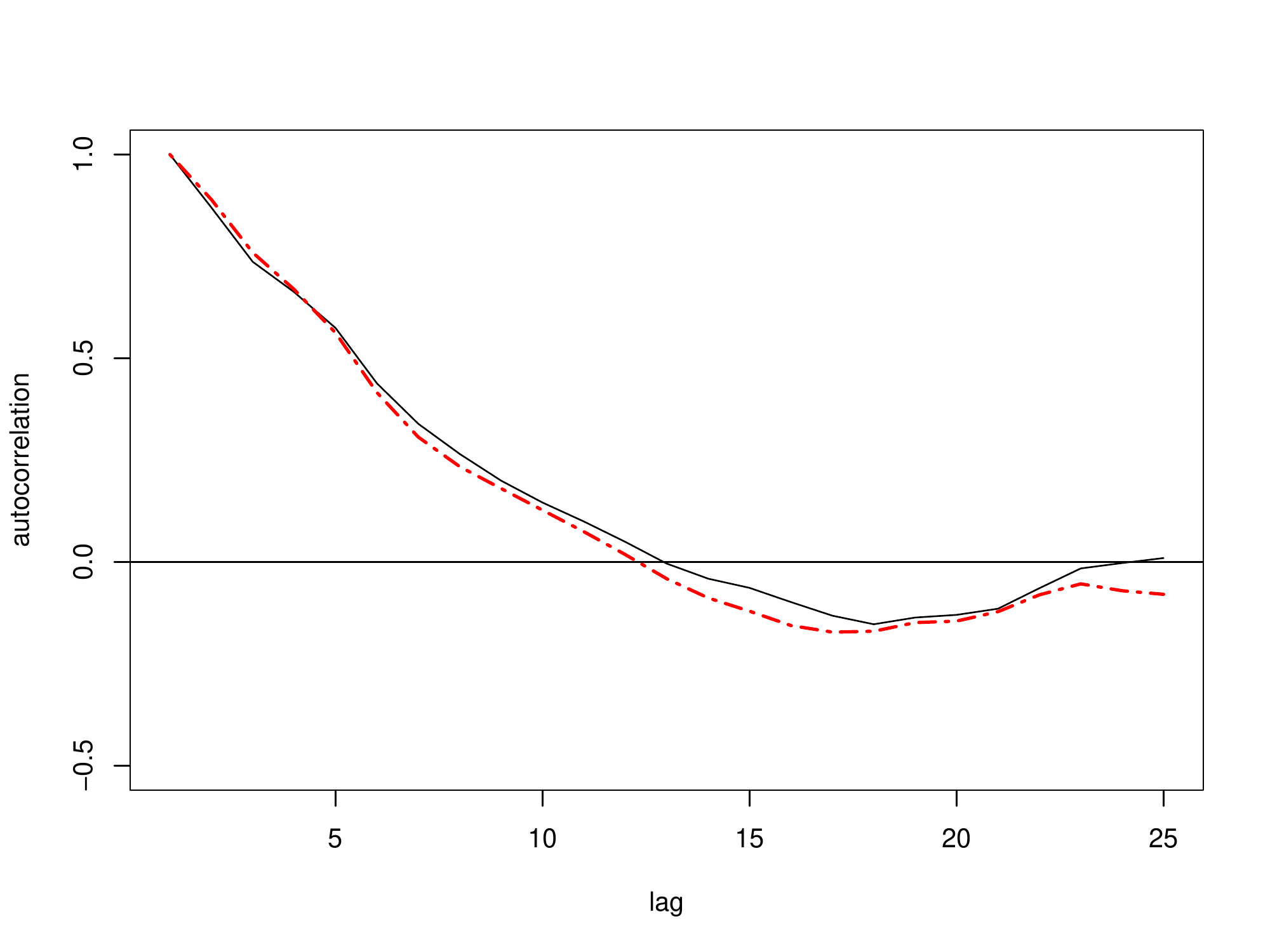}
\caption{Original series (black) and privatized version (red).  Top left: Sample paths of the detrended series for the case of low cross-correlation. Top right: Sample 
autocorrelation plots for the case of low cross-correlation.  Bottom left:  Sample paths of the detrended series for the case of high cross-correlation. Bottom right: Sample autocorrelation plots for the case of high cross-correlation.  }
\lb{fig:example1}
\end{figure}

Figure~\ref{fig:prvcy_util} shows the aggregate results for the 500 Monte Carlo replications. The average value (over 500 replications) of the sample version of the privacy measure \eqref{eq:prvcy} was greater than 0.99 for both values of the cross-correlation between $X_t$ and $Z_t$. 
To look at the pathwise difference between the sensitive series and the privatized series we used a normalized squared distance: 
\[    D_{path} = \frac{T^{-1}\sum_{t=1}^T (X_t - {\hat{X}}_t)^2}{\gamma(0)}, \]
where $\gamma(0) = \mbox{Var} (X_t).$  This measures
the average pointwise discrepancy between the
original time series and its privatized version,
relative to the process' variation.  Figure~\ref{fig:prvcy_util} shows the distribution of $D_{path}$ over the 500 replications. The  pointwise difference between $X_t$ and ${\hat{X}}_t$ is larger than one standard deviation of $X_t$ for at least 50\% of the simulations, and is bigger than $80\% $ of the standard deviation in more than 60\% of the simulations. This is true for both values of the cross-correlation. 

We also examined the squared distance between the sample autocorrelations of the original series and the FLIP-ped series, measured as 
\[ 
D_{ACF} = H^{-1}\sum_{h=0}^H (\rho_h - {\hat{\rho}}_h)^2, 
\] 
where $\rho_h$ and ${\hat{\rho}}_h$ are the lag $h$ sample autocorrelations for $\{ X_t \}$ and $\{ {\hat{X}}_t \}$, respectively.  For the reported results we set $H = 24$.  Figure~\ref{fig:prvcy_util} shows the distribution of $D_{ACF}$ over the 500 replications for the two different cross-correlation setting. In general, the two autocorrelations are close to each other, indicating high utility of the FLIP-ped series. \\

\begin{figure}[htbp!]
\centering

\includegraphics[width = 1.8in, height = 1.2in]{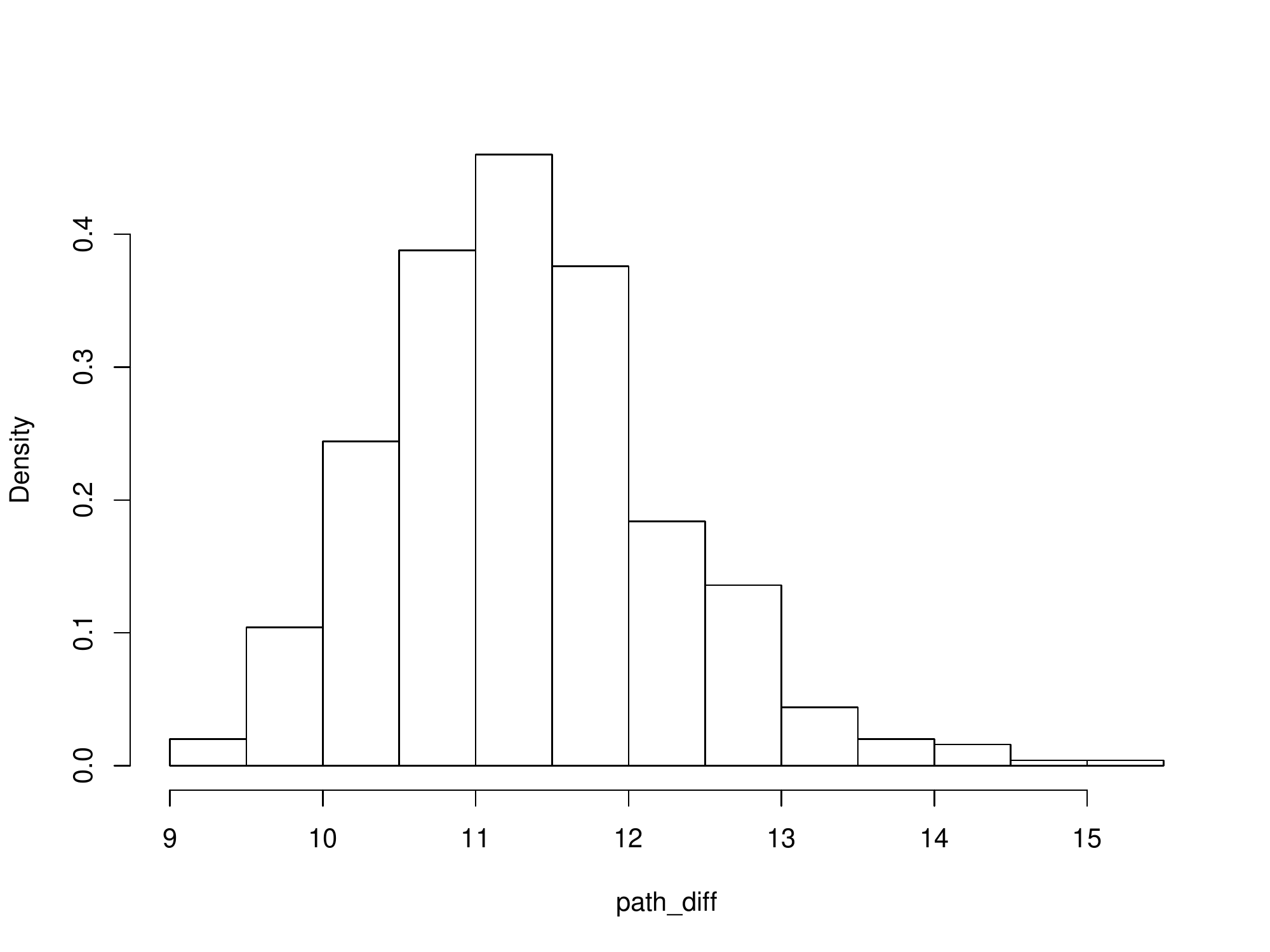}
\includegraphics[width = 1.8in, height = 1.2in]{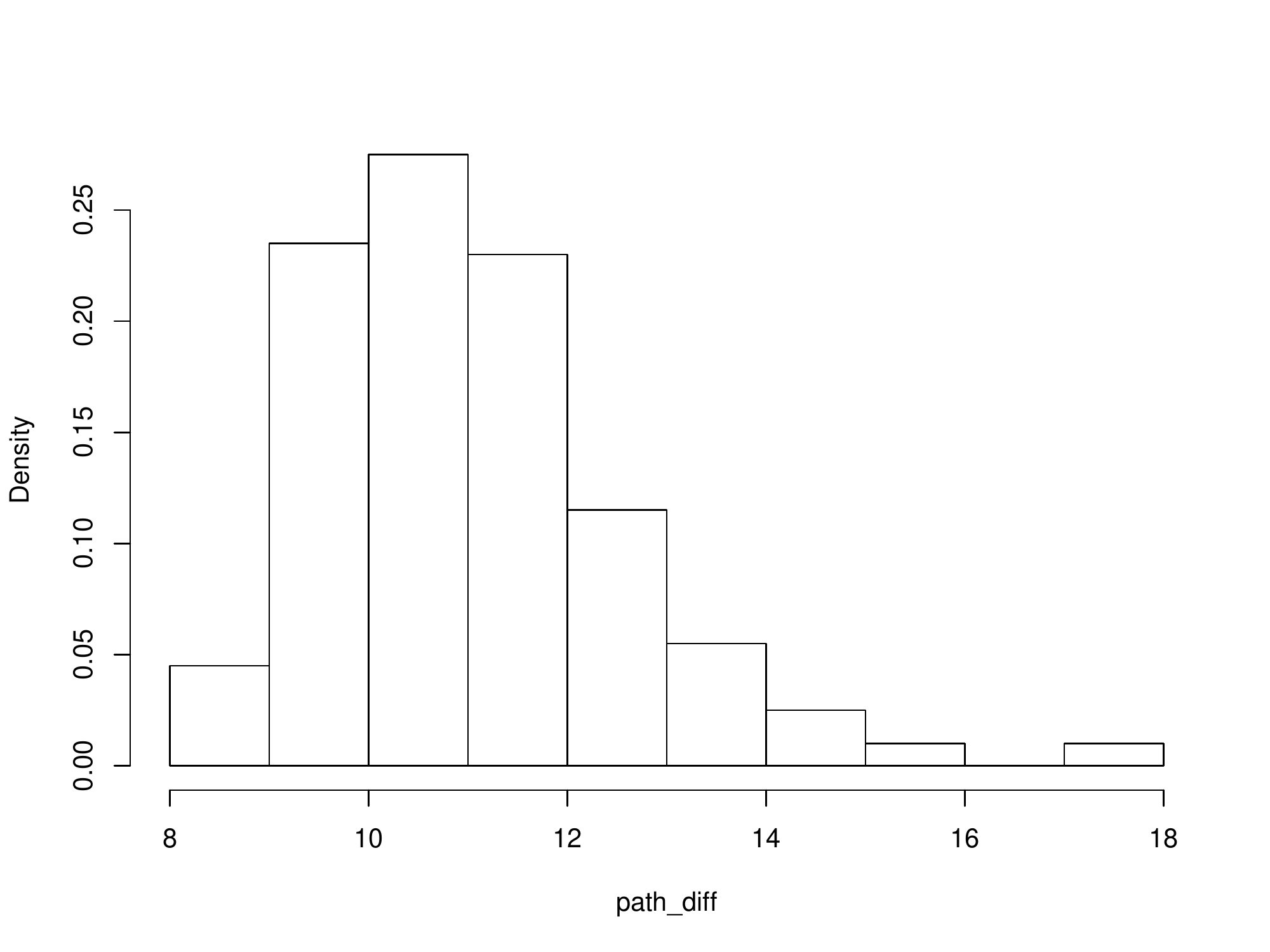}\\
\includegraphics[width = 1.8in, height = 1.2in]{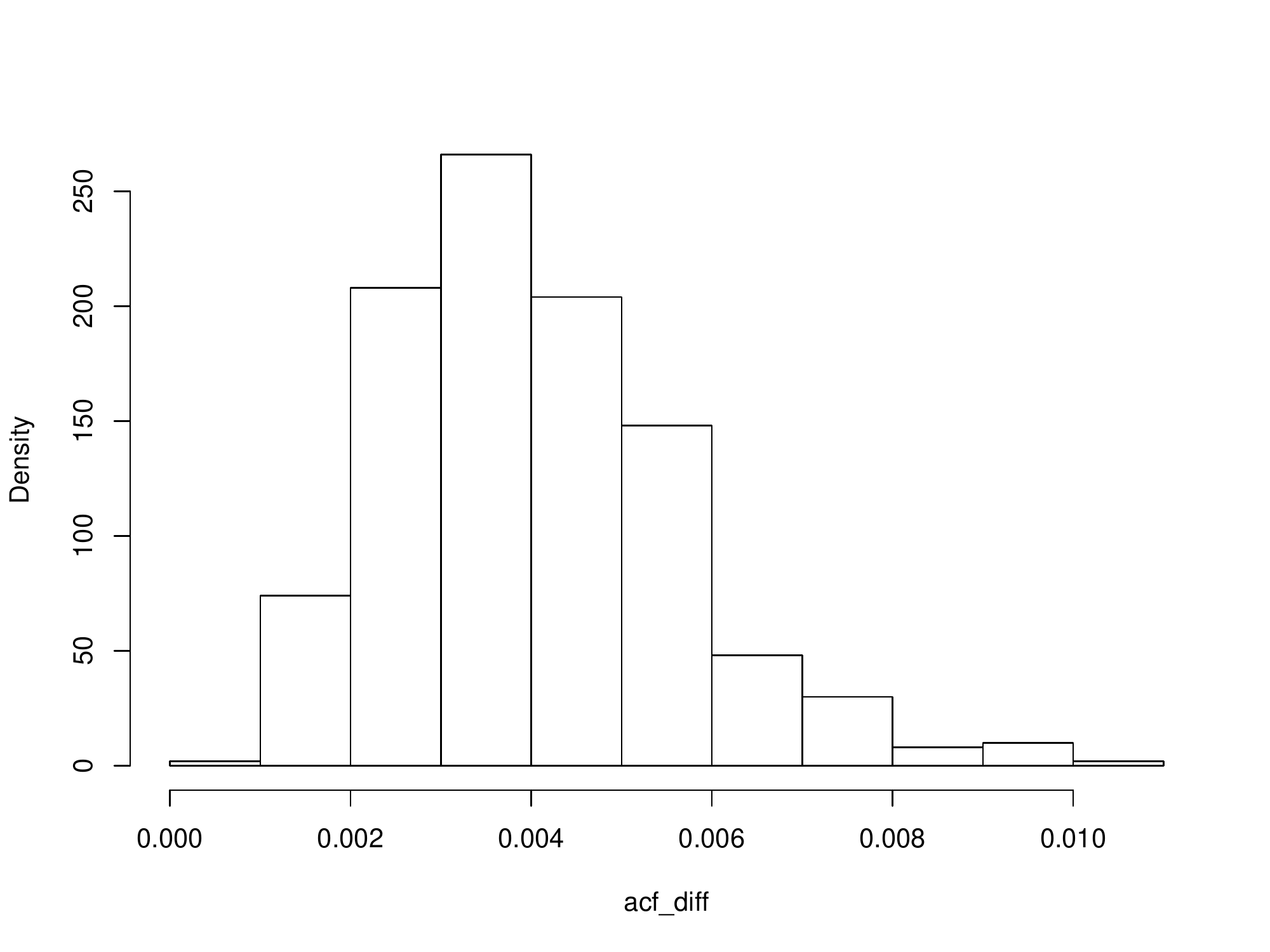}
\includegraphics[width = 1.8in, height = 1.2in]{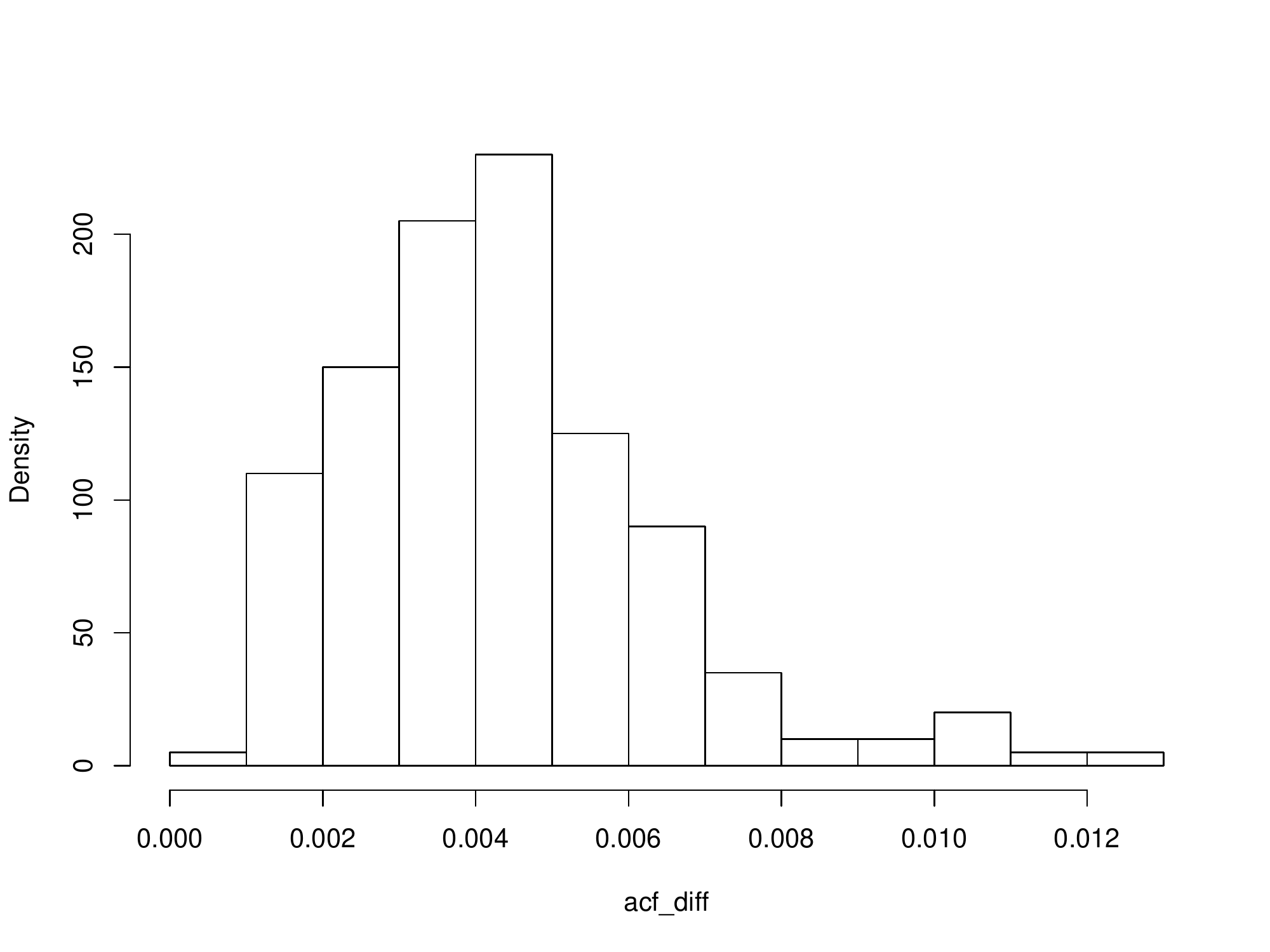}
\caption{Top left: $D_{path}$ histogram over
$500$ simulations for the case of low ($\rho=0.1$) cross-correlation.  Top right: $D_{path}$ histogram over $500$ simulations for the case of high ($\rho=0.7$) cross-correlation. Bottom left: $D_{ACF}$ histogram over $500$ simulations for the case of low ($\rho=0.1$) cross-correlation.  Bottom right: $D_{ACF}$ histogram over $500$ simulations for the case of high ($\rho=0.7$) cross-correlation. }
\lb{fig:prvcy_util}
\end{figure}
}}
\end{example}

\begin{example}
\lb{exmp:example2}
\noindent{\underline{\bf $\delta-$FLIP mechanism for $\delta > 0.$}}\\
{\rm{
In the next example we consider a model with a linear trend, and demonstrate the usefulness of the $\delta-$FLIP algorithm in leaving the trend unchanged while perturbing the stationary disturbance around the trend. 
We consider the model \eqref{eq:obs_model}, where the stationary part $\{ X_t, Z_t\}$ is assumed to follow the model \eqref{eq:obs_model} from Example~\ref{exmp:example1}. The VAR(1) coefficient matrix and the error variance matrix are chosen to be as in Example~\ref{exmp:example1}. The linear trends for the sensitive series and the auxiliary series are chosen to be 
\beqa
\mu^X_t &=& 30 + 0.05 t, \\
\mu^Z_t &=& 10 + 0.06 t.
\eeqa
The privacy budget $\delta$ is chosen to be 0.1. The data is detrended using ordinary least squares, by regressing each series $\{ X_t \}$ and $\{ Z_t \}$ separately on $\{ 1, t \}$. Then the FLIP algorithm is applied to the least squares residual series to obtain an all-pass filter that will leave trend of order $d =1$ unchanged. The estimated trend then is added back to the filtered least squares residuals to obtain the final privatized series. 

The left panel of Figure~\ref{fig:example2} shows the original sensitive series (solid black) along with the filtered version (dashed red) for a typical series generated using the assumed model. The middle panel shows the least squares residual series (solid black) along with the perturbed series (dashed red) after application of a $\delta-$FLIP filter. 
The right panel shows the sample autocorrelation of the least squares residual series (solid black) along with that of the perturbed series (dashed red). 
The results of the simulation on the least squares residuals were very similar to those obtained in Example
\ref{exmp:example1}  using the stationary series, and hence are not reported here.

\begin{figure}[htbp!]
\centering

\includegraphics[width = 1.8in, height = 1.2in]{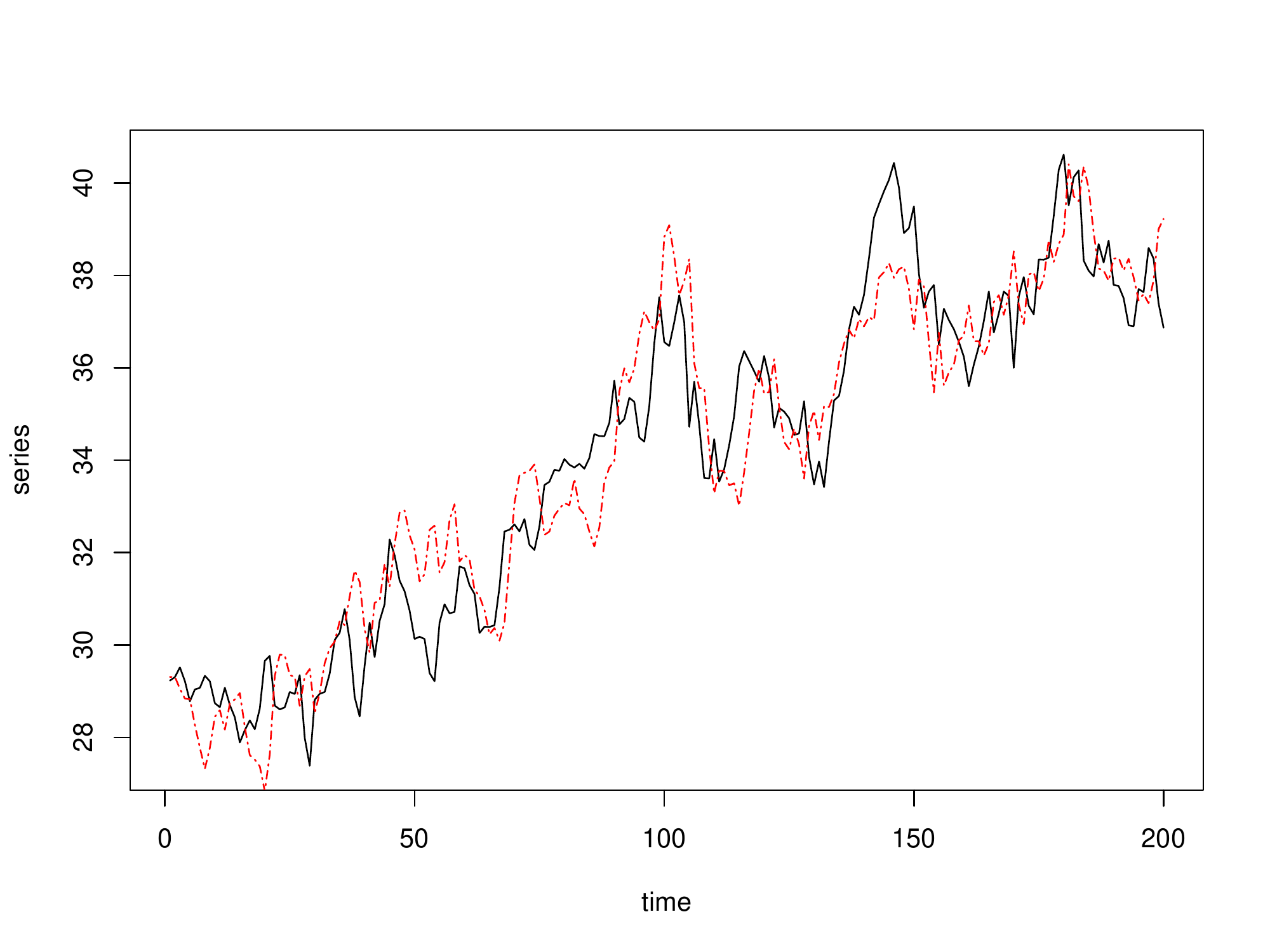}
\includegraphics[width = 1.8in, height = 1.2in]{VAR1_ex2.pdf}
\includegraphics[width = 1.8in, height = 1.2in]{VAR1_ex2_autocorr.pdf}
\caption{Original series (black) along with privatized version (red) for a linear trend model.  Left: Sample paths with trend.  Middle: Sample paths without trend.
Right: Sample autocorrelation plots for both series.}
\lb{fig:example2}
\end{figure}
}}
\end{example}

\begin{example}
\label{QWI data}
\noindent{\underline{\bf  FLIP-ping Quarterly Workforce Indicators.}} \\
{\rm{
To illustrate how the proposed FLIP methodology can be used on a regularly sampled time series that is published by a statistical agency, we examine
Quarterly Workforce Indicator (QWI) data published by the U.S. Census Bureau.  The indicators are based on different jobs and work location administrative data from 49 states, and are available quarterly; see 
\cite{abowd2011national} for more details about the construction and publication of the data.
All data were extracted from the QWI Explorer website \cite{QWI} on 17th March 2022 at 10:45 pm. We focus upon the quarterly indicator ``Beginning of Quarter Employment: Count" (or {\em employment count} for short)
for the states of California and Maryland, with an
observation period of Q1 1995 through Q4 2019.

The QWI are available at different levels of aggregation, and the granularity of finer cross-tabulation of these data -- based on factors such as geographic region, age, race, sex and education category -- may lead to disclosure of sensitive information on demographic and economic details of local labor markets. While higher level aggregation are less susceptible to disclosure risk, finer partitions of the data need to be protected. Below, we illustrate how the FLIP mechanism can be used to perturb lower level aggregation of the QWI series while retaining important time series features, so as to make the public use data dynamically consistent. We use published data, and hence there is no risk of disclosure. Thus, the illustration is essentially a data-driven example constructed for illustrative purposes.

\noindent {\bf 1. California Employment Count Across Races:} 

Here we examine the QWI employment count for California, where the data is cross-tabulated by race. No race or ethnic group constitutes a pure majority of California’s population: 39\% of state residents are Latino, 35\% are white, 15\% are Asian American or Pacific Islander, 5\% are Black, 4\% are multiracial, and fewer than 1\% are Native American or Alaska Natives, according to the 2020 Census. We choose quarterly employment count of whites as the series known to the attacker, i.e.,  the $\{ Z_t \}$ series in our proposal. The count of the Asian group is chosen  as the series $\{ X_t \}$ that needs to be protected from attackers. 
 Below, we report the results of applying the $\delta-$FLIP mechanism to the employment count data for $\delta = 0$ and   $\delta = 0.1.$
 
 \begin{figure}[htbp!]
\centering

\includegraphics[width = 1.8in, height = 1.2in]{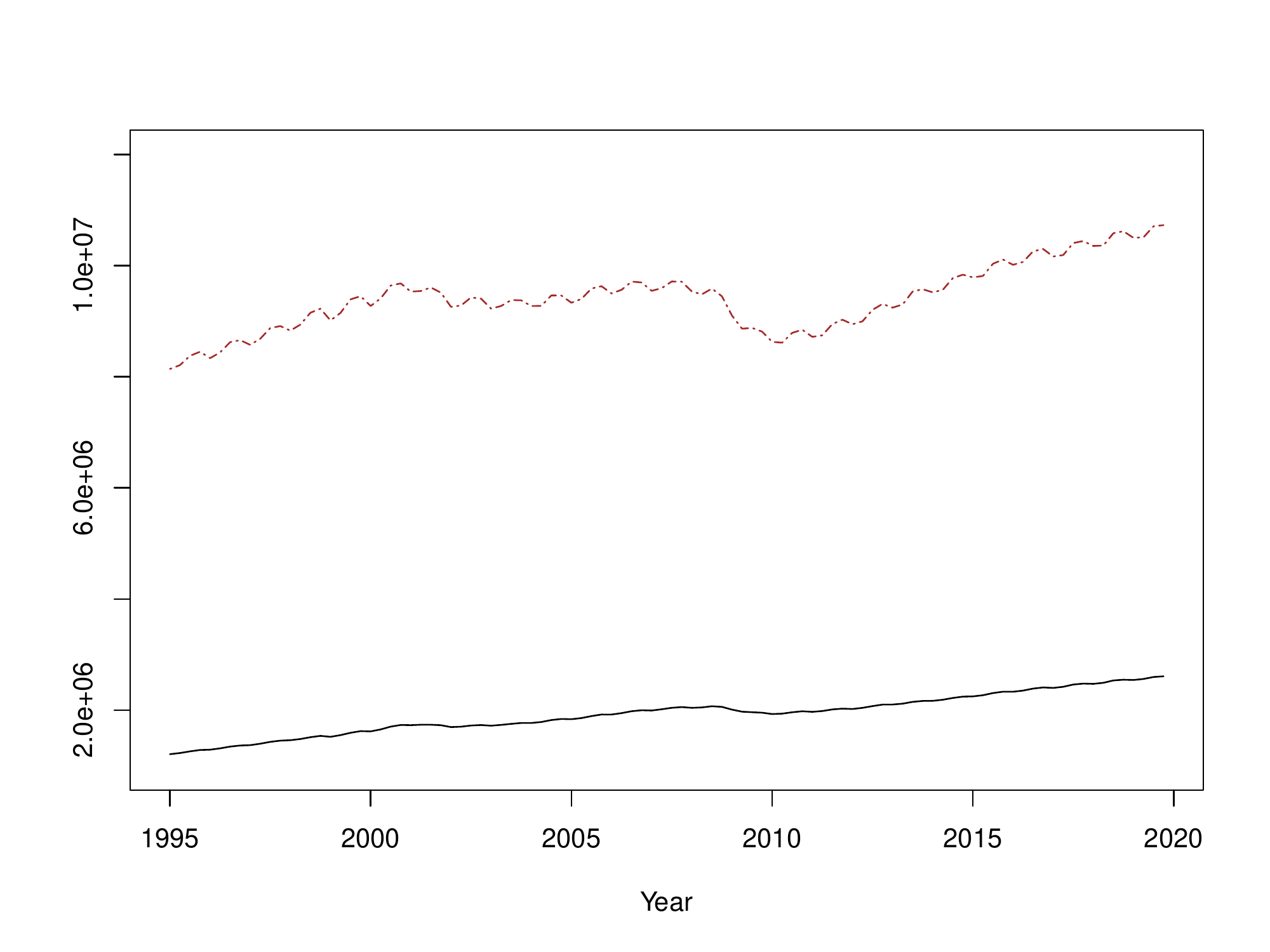}
\includegraphics[width = 1.8in, height = 1.2in]{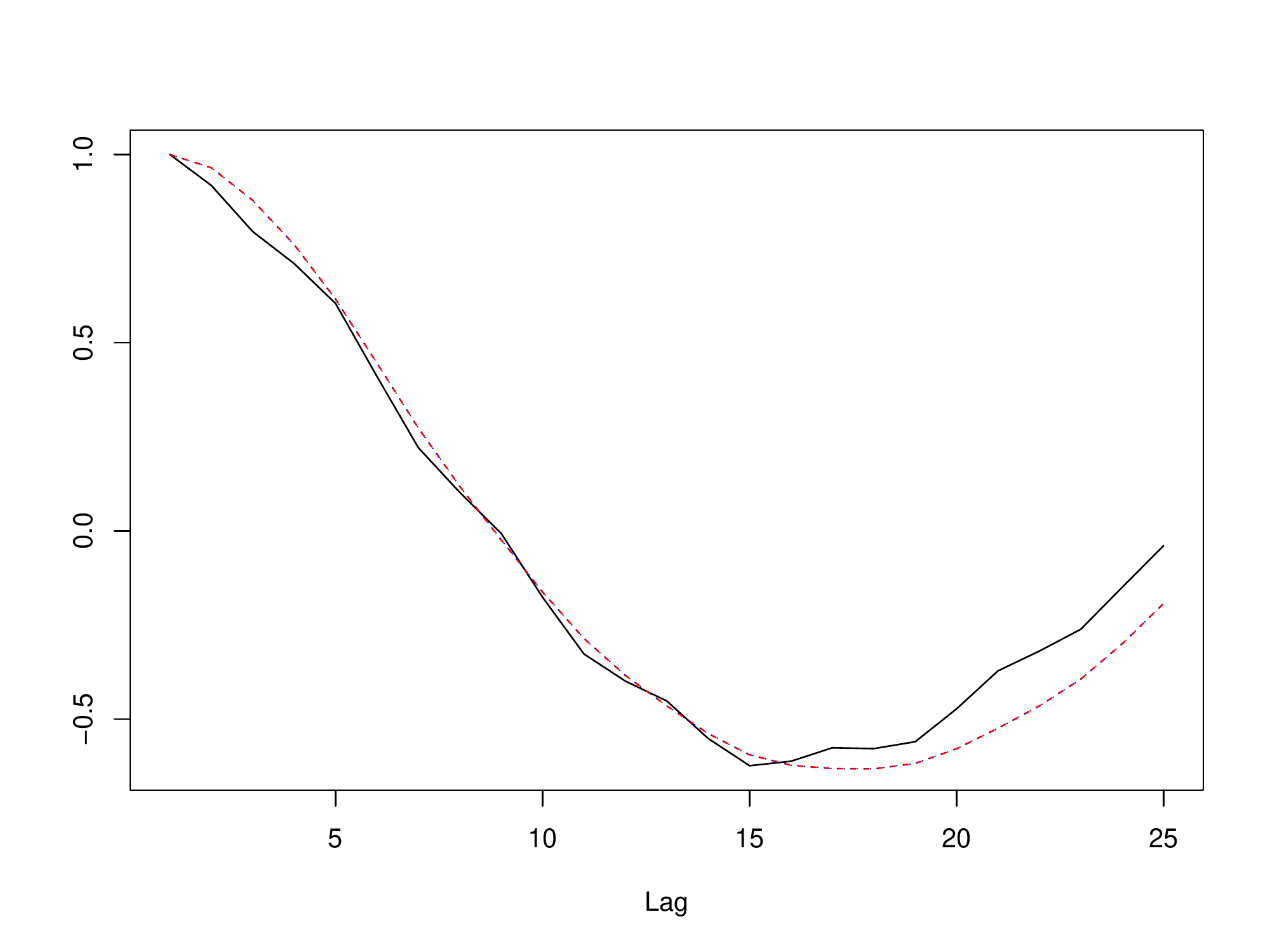}\\
\includegraphics[width = 1.8in, height = 1.2in]{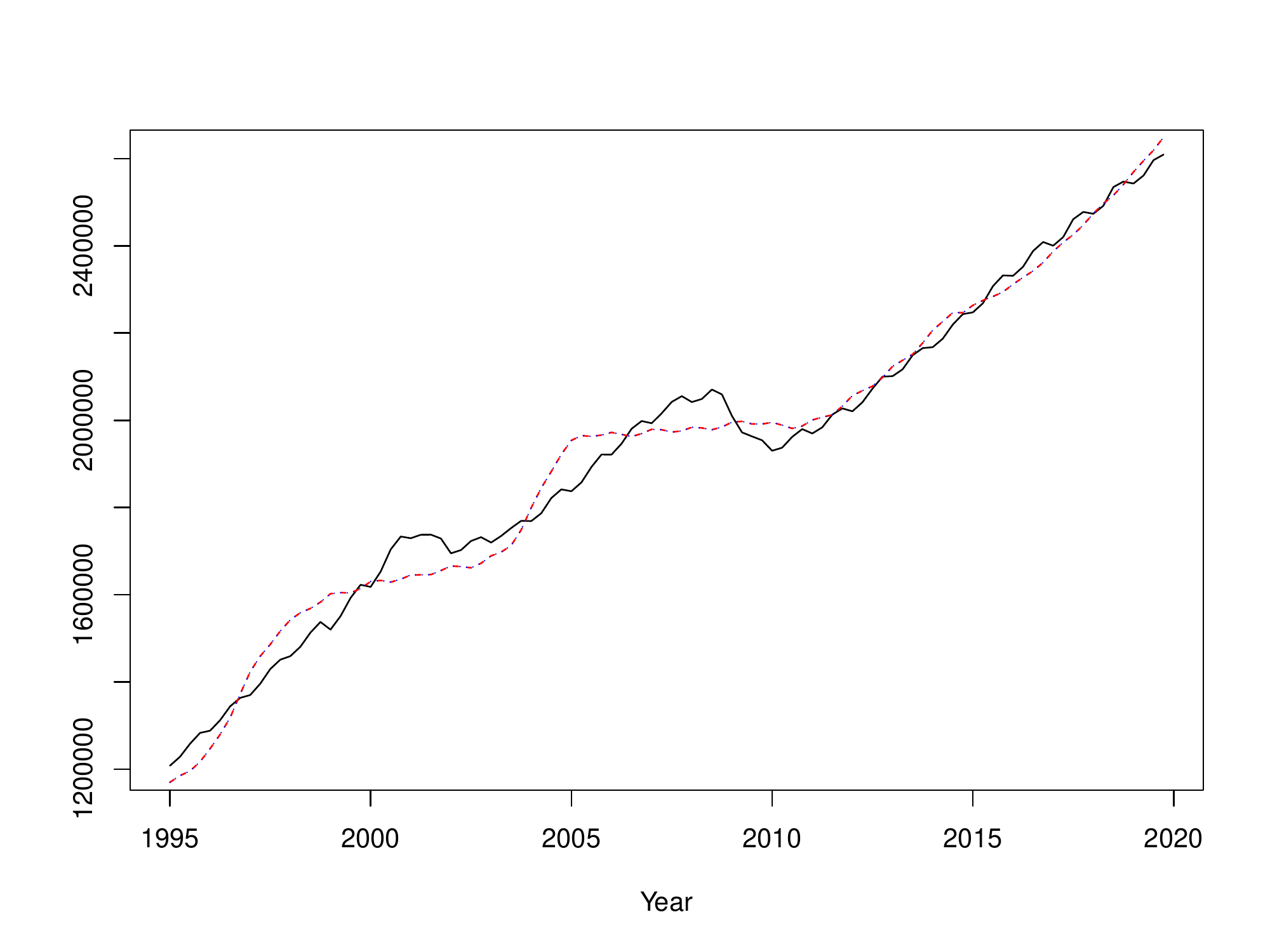}
\includegraphics[width = 1.8in, height = 1.2in]{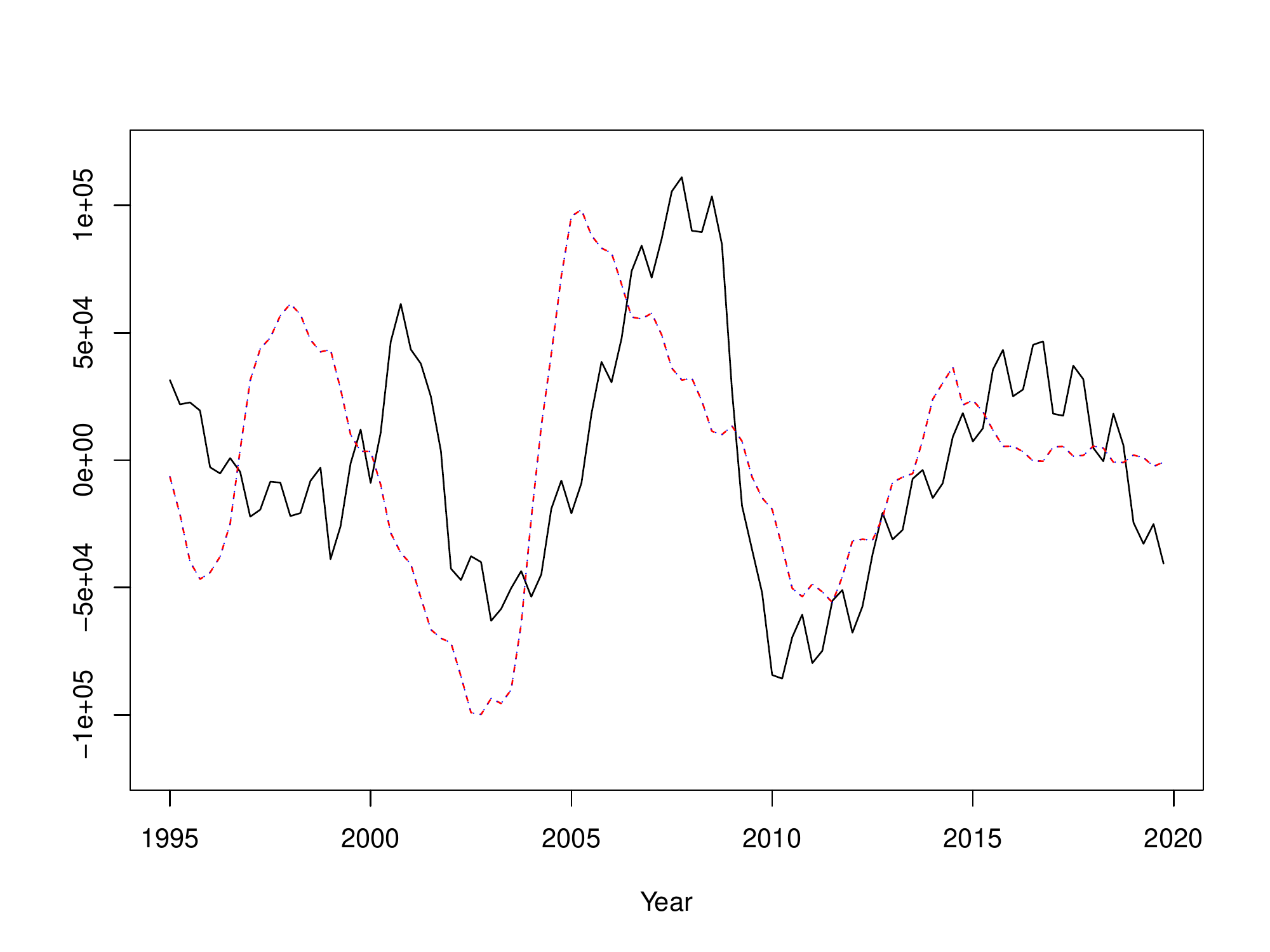}
\caption{Top left: Employment counts for Asians (black) and Whites (brown) in California. Top right: Sample autocorrelation for  Asian employment count (black) and privatized versions ($\delta = 0$ in red and $\delta = 0.1$ in blue). Bottom left: Asian employment count (black) and privatized versions ($\delta = 0$ in red and $\delta = 0.1$ in blue). Bottom right: Detrended Asian employment count (black)  and  privatized versions ($\delta = 0$ in red and $\delta = 0.1$ in blue).
Note: The graph with the two series are shown in their original scale.}
\lb{fig:example3}
\end{figure}

\noindent
\underline{\textbf{Case: $\delta=0:$}} \label{case1} The values of the two series are in millions, but for implementation of the FLIP methodology we work with the standardized version of both the series,
obtained by subtracting the sample mean and  dividing by the sample standard deviation. 
Because a linear trend seems to be inadequate
to capture the stable growth pattern for both 
$\{ X_t \}$ and $\{ Z_t \}$, we assume $d=3$, i.e. the trend is a polynomial of order three. 
As in Example~\ref{exmp:example2}, we use the residuals from the trend fit as the $\{ X_t \}$ and $\{ Z_t \}$  series; also we set $K=25$ and $M=25$.
The beta mixture in the definition of the $R$ function uses two components. The measures $D_{path}$ and $D_{ACF}$ are calculated as described in Example~\ref{exmp:example1}, yielding 
 $0.9549$ and $0.0016$ respectively. A way to interpret the measures would that on an average the difference between the original time series and the perturbed one is expected to be about 97.7\% ($\sqrt{0.9545} = 0.977$) of the  standard deviation of the original series and the autocorrelations for each lag of the perturbed series are expected to be on an average within $\pm 0.04$ of the original sample autocorrelations. The privacy measure evaluated for the series  is $0.9988$; 
results are plotted in Figure~\ref{fig:example3}. As intended, there are substantial differences between the stationary parts of the observed series and the perturbed series, whereas the sample autocorrelation estimates are nearly identical. 

\noindent
\underline{\textbf{Case: $\delta \neq 0:$}}  We repeat the above procedure ($\delta=0$), but now using $\delta = 0.1.$ The $D_{path}$ in this example is $0.9328$, and $D_{ACF}$ is $0.0026$; the privacy measure is $0.9982.$ Figure ~\ref{fig:example3} show the original series and the perturbed series along with their sample autocorrelation values.
 The top left panel shows the two series together (Asians in red and Whites in black). In the rest of the panels results related to the original series for Asian employment counts is shown in black, whereas those related to the two privatized versions for $\delta = 0$ and $\delta =0.1$ are shown in red and blue, respectively. The top right panel shows the sample autocorrelation of the original and the two privatized series. The bottom left panel shows the Asian employment count series along with the two privatized series, whereas the bottom right panel shows those for the detrended version. 
In the case of $\delta = 0.1$ the perturbations  are somewhat -- though not substantially -- more than those in the $\delta = 0$ case, and the sample ACF of the perturbed series is more distinct from that of the original series.

\noindent {\bf2. Maryland Employment Count Across Counties:}

We repeat the exercise with the Maryland employment count QWI, examined across different counties. We choose Montgomery county as the $\{ Z_t \}$ series, keeping Baltimore county as the $\{ X_t \}$ series; again we use a third order polynomial trend.  In Figure~\ref{fig:example3partMD1} we plot the sample autocorrelations, the time series sample paths, and the detrended series for both the original data and the privatized ($\delta=0$ and $\delta=0.1$) versions. The top left panel shows the two series together (Baltimore county in black and Montgomery county in red). There is substantial contemporaneous correlation between the two series. That explains the fact that the perturbation amounts for $\delta=0$ and $\delta=0.1$ are not very different, resulting in  two  almost identical privatized series. In the remaining panels results related to the original series for Baltimore county are shown in black, whereas those related to the two privatized versions for $\delta = 0$ and $\delta =0.1$ are shown in red and blue, respectively. The top right panel shows the sample autocorrelation of the original and the two privatized series. The bottom left panel shows the original series along with the two privatized series, whereas the bottom right panel shows the sample autocorrelation for the orginal Baltimore county series and for two privatized versions ($\delta = 0$ in red and  $\delta = 0.1$ in blue).

\begin{figure}[htbp!]
\centering

\includegraphics[width = 1.8in, height = 1.2in]{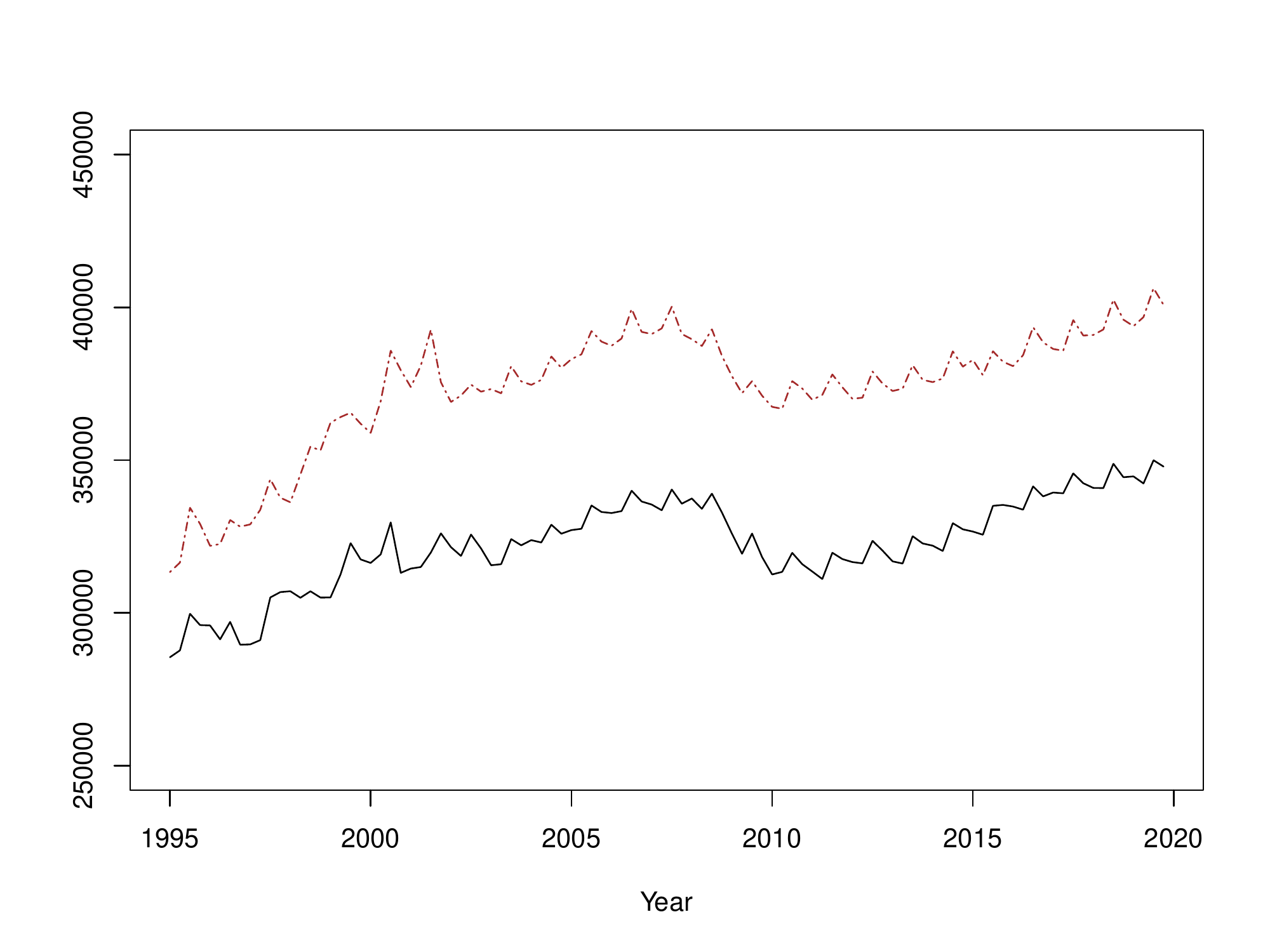}
\includegraphics[width = 1.8in, height = 1.2in]{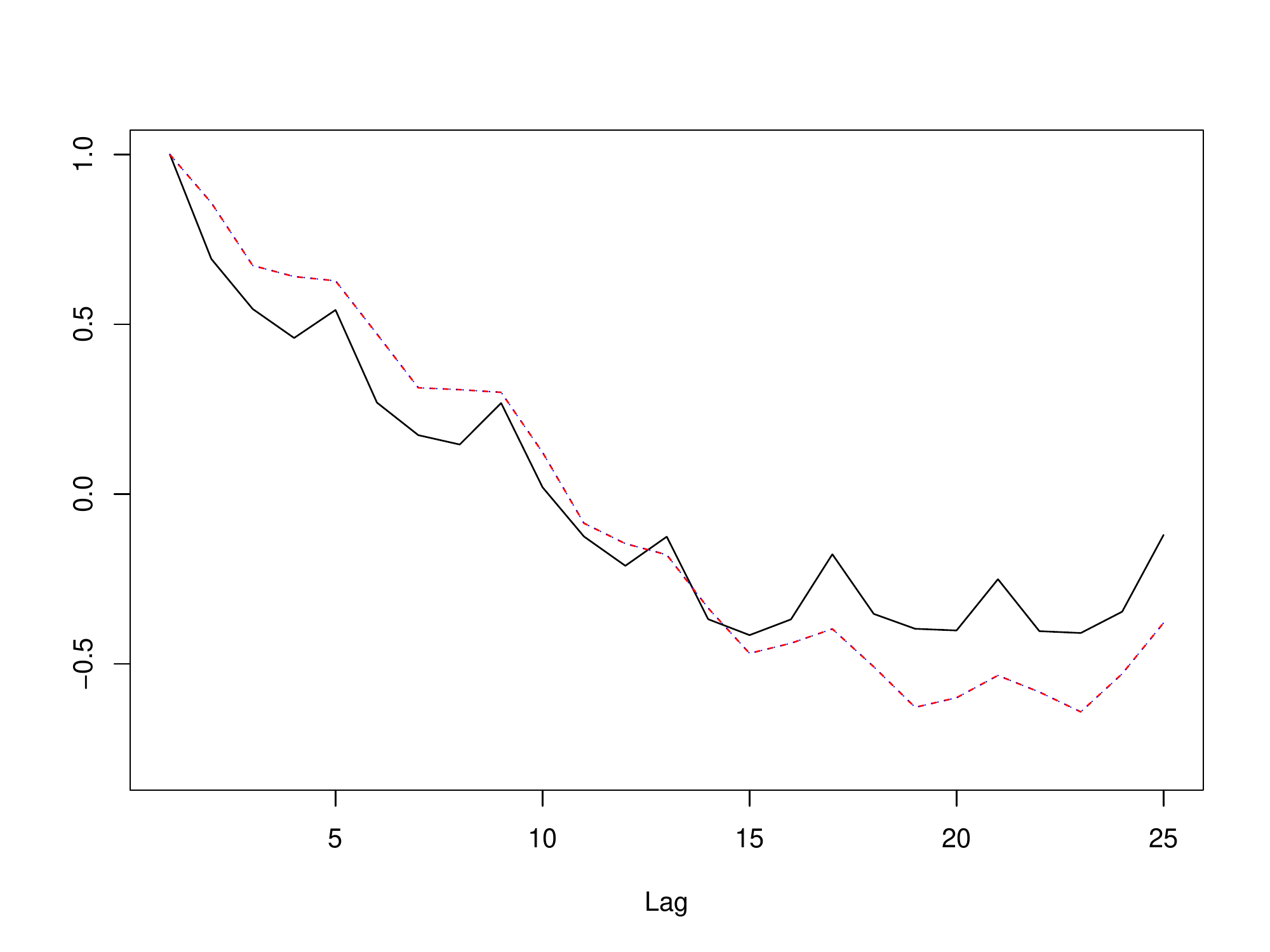}\\
\includegraphics[width = 1.8in, height = 1.2in]{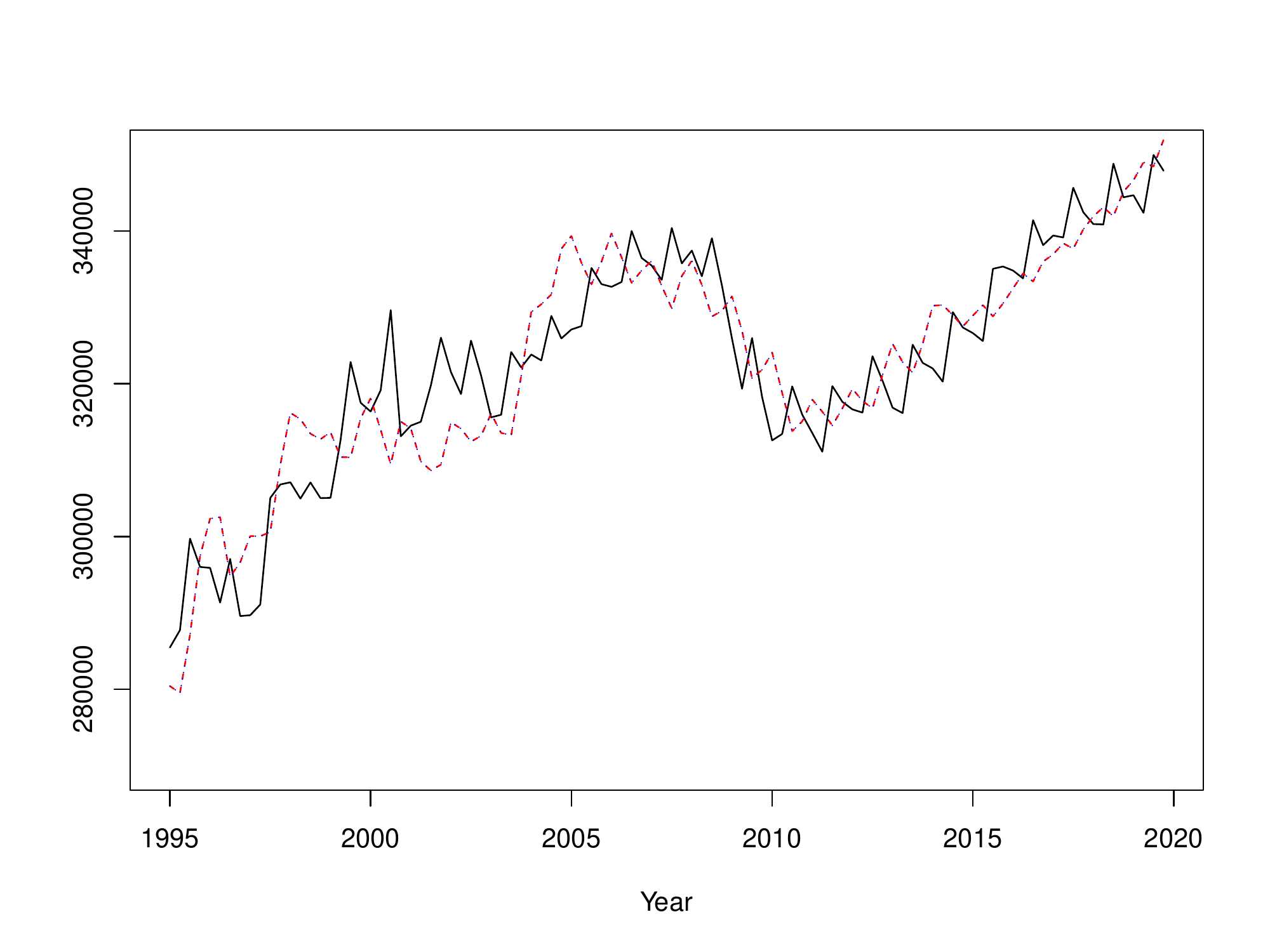}
\includegraphics[width = 1.8in, height = 1.2in]{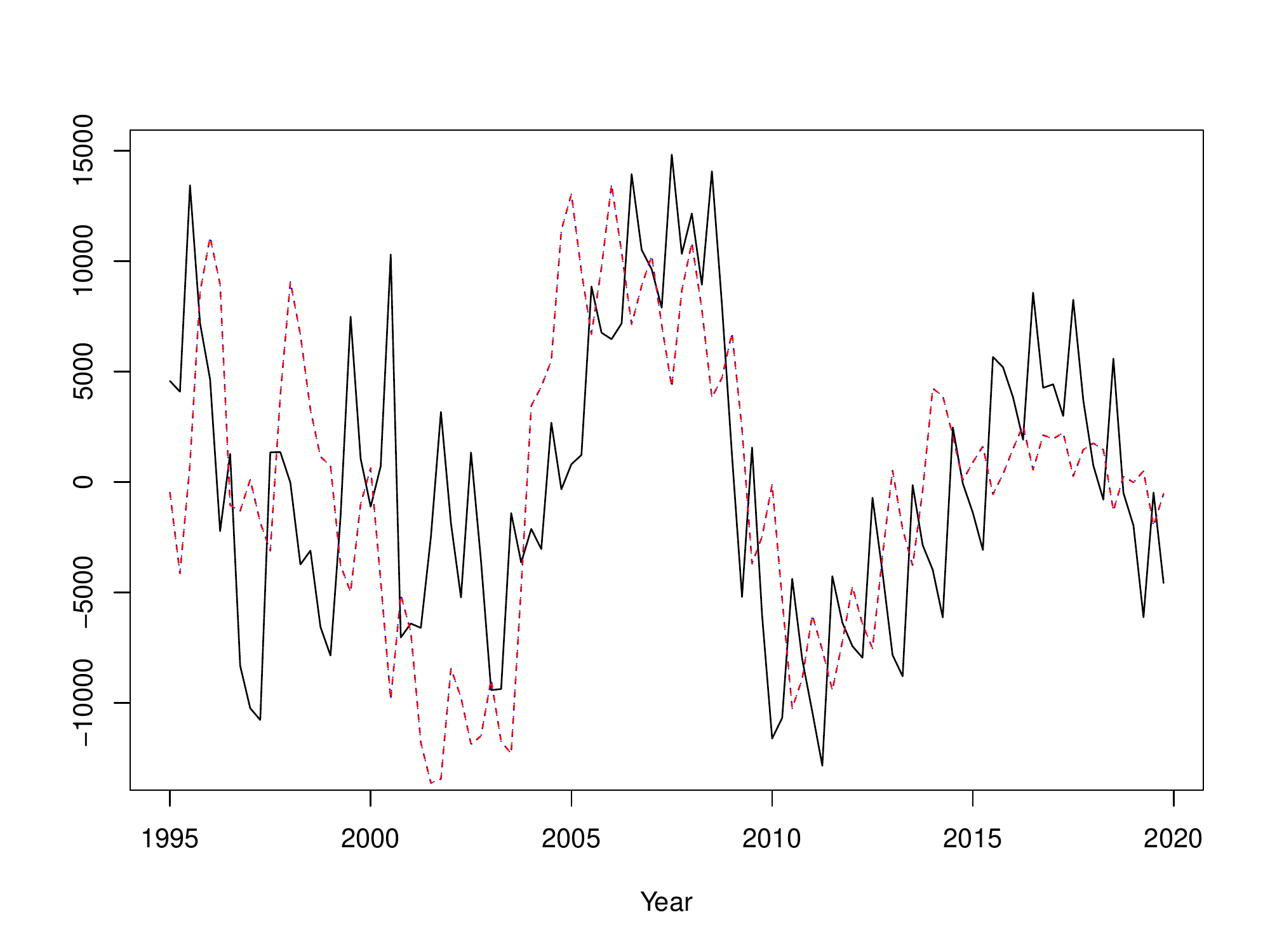}
\caption{Top left: Employment counts for Baltimore county (black) and Montgomery county (red) in Maryland. Top right: Sample autocorrelation for original Baltimore county employment count (black) and privatized versions ($\delta = 0$ in red and $\delta = 0.1$ in blue). Bottom left: Baltimore county employment count (black) and privatized versions ($\delta = 0$ in red and $\delta = 0.1$ in blue). Bottom right: Detrended Baltimore county employment count (black) and  privatized versions ($\delta = 0$ in red and  $\delta = 0.1$ in blue). Note: The graph with the two series are shown in their original scale.}
\lb{fig:example3partMD1}
\end{figure}
}}
\end{example}

\section{Conclusion}
We have proposed a framework for disclosure avoidance of time series data. The methodology is particularly suitable for regularly spaced time series data where the data publisher anticipates the amount of sensitive information that may be available to the attacker in the form of an auxiliary time series that is cross-correlated with the sensitive time series.  Specifically, we demonstrate that
\begin{itemize}
    \item Noise addition alters serial correlation structure, and thereby mars utility
    \item All-pass filtering preserves the second-order structure of a time series, and hence maintains certain forms of utility
    \item Linear Incremental Privacy (LIP) can assess privacy protection for stationary time series data
    \item Feasible LIP (or FLIP), which uses all-pass filtering with a specified privacy budget $\delta \geq 0$, is a mechanism that can protect time series data.
\end{itemize}
The novelty of the FLIP proposal consists in how privacy protection is measured (as the inability to improve the attacker's knowledge), as well as 
providing -- unlike conventional randomized privacy mechanisms -- full analytical validity, in terms of preserving the sample path's dependence structure.

Although our focus is on stationary time series,
 we have provided extensions to allow for non-stationary trend effects through the all-pass filter design.  We remark that here we have focused on first and second order structure of a time series, i.e., the mean function and serial correlation pattern; it is natural to consider higher order structures of the time series, possibly assessed through higher order polyspectra, and investigate an analogue of all-pass filter that
 may also preserve such properties.  In future work we plan to research further extensions to non-stationarity and higher-order dynamics.

\newpage
\appendix
\section*{Appendix A.}
\noindent{\bf Proof of Proposition~\ref{prop:prvcy_measure}}
\begin{proof}
Let the lag $j$ autocovariance matrix for
the multivariate process $\{ X_t, Z_t \}$
be denoted
\[
\Gamma_j= \begin{pmatrix}
\gamma_X(j) & \gamma_{XZ}(j) \\
\gamma_{XZ}(-j) & \gamma_Z(j)
\end{pmatrix}, 
\]
where $\gamma_{XZ}(j)=E(X_{t+j}Z_t)$ and $\gamma_{ZX}(j)=E(Z_{t+j}X_t)=\gamma_{XZ}(-j).$
 The corresponding cross-spectral densities will
 be denoted by $f_{XZ}$ and $f_{ZX}$.
The linear projection of $X_t$ on $\{ Z_t \}$ is 
given by 
\[ 
   E(X_t| \{ Z_t \} )= \sum_k \pi_k Z_{t-k}
\]
 for some coefficients $\{ \pi_k \}$ to be determined.
This is the best guess (in the sense of
minimum MSE among linear predictors) 
of the original time series at time
point $t$ made by the attacker based on his knowledge of $\{Z_t\}$, and from the normal equations it follows
that
\[
\gamma_{XZ} (h) = E (X_t Z_{t-h}) = 
  E (  E [X_t| \{ Z_t \} ]  Z_{t-h})
 = \sum_k \pi_k \gamma_Z (h-k)
\]
for any integer $h$.  Taking the Fourier transform
 yields $f_{XZ} (\lambda) = \pi (e^{-i \lambda}) f_Z (\lambda)$, and hence
 \begin{equation}
     \label{eq:pi-predictor}
  \pi (e^{-i \lambda}) = f_{XZ} (\lambda) /f_Z (\lambda).
  \end{equation}
 Next, the autocovariance function of 
 $X_t - E(X_t \vert \{Z_t\})$ is 
\begin{eqnarray*}
\mbox{Cov}(X_t,X_{t-j}| \{ Z_t \} )
 &=&\mbox{Cov}(X_t-E(X_t| \{ Z_t \} ),
   X_{t-j}-E(X_{t-j}| \{ Z_t \}) ) \\ 
&=&\gamma_X(j)-\sum_l \pi_l \gamma_{XZ}(l+j)-\sum_k \pi_k \gamma_{XZ}(k-j)\\
&& +\sum_k \sum_l \pi_k \pi_l \gamma_Z(j+l-k)
\end{eqnarray*}
for any integer $j$, and taking the Fourier transform
yields (\ref{eq:cond_spec}) 
via (\ref{eq:pi-predictor}).
 Similar calculations yield
\[
\mbox{Cov}(X_t, \hat{X}_{t}| \{ Z_t \} )
 = \langle \Psi, f_{X \vert Z} \rangle \quad
 \mbox{and} \quad
 \mbox{Cov}(\hat{X}_t, \hat{X}_{t}| \{ Z_t \} )
 = \langle \Psi \overline{\Psi}, f_{X \vert Z} \rangle,
 \]
 and the formula for the privacy measure follows
 so long as $f_{X \vert Z} > 0$.

\end{proof} 

\noindent{\bf Proof of Theorem~\ref{thm:perfect_prvcy}}
 \begin{proof}
 Let $\Psi(z)= \exp \{ \phi(\lambda) \} = 
 \exp \{ \sum_k \phi_k z^k \}$ be the cepstral representation \eqref{eq:all-pass} of $\Psi$. 
Since $|\Psi(z)| = 1$ for all $z = e^{-i \lambda}$, we have $\phi_k = -\phi_{-k}$ and $\Psi(e^{-i\lambda}) = \exp \{ i g(\lambda) \}$, where $ g(\lambda) = - 2 \sum_{k \geq 1} \phi_k  \sin (\lambda k).$   Thus we want to find odd functions $g$ with expansion  $-2 \sum_{k \geq 1} \phi_k  \sin (\lambda k)$ that will provide a perfect privacy solution. 
For such an odd function, perfect privacy  would necessarily imply $\langle \cos(g), f\rangle_{\pi} = 0.$
Now assume that $g$ has the form 
 $g(\lambda) = \pi R(F(\lambda))$for $\lambda \in [0, \pi],$ where $F(\lambda) = \int_0^{\lambda} \tilde{f} (\omega) d\omega$ and $R:[0, 1] \to [0, 1]$. We need to show that an $R$ function with  $\langle \cos(\pi R(F), f\rangle_{\pi} = 0$ exists. Consider any
$R \in    \mathcal{R}_{f}$. 
By the change of variable $x = F(\lambda)$,
$\langle \cos(\pi R(F)), f\rangle_{\pi} =\int_0^1 \cos(\pi R(x)) dx.$ 
Since $R(x) = 1 -  R(1 - x) $ and $\cos(\theta) = -\cos(\pi - \theta)$ for $\theta \in [0, \pi]$, we have $\langle \cos(\pi R(F)),  f\rangle_{\pi} = 0.$
Extending $R$ to [-1, 0] via $R(-x) = - R(x)$, and choosing $g(\lambda) = \pi R(F(\lambda))$ for $\lambda \in [-\pi, \pi]$, we obtain the result. 
\end{proof}

\noindent{\bf Proof of Theorem~\ref{thm:deltaLIP}}
\begin{proof}
Let $h$ be a spectral density belonging to $\mathcal{F}_R(f, \delta)$ for some $R \in \mathcal{R}$.  Then  for $z = e^{-i \lambda}$
\beqa
|\Psi_h(z) - \Psi_f(z)| &=& | \exp \{i \pi R(H(\lambda)) \} - \exp \{i \pi R(F(\lambda) \}| \\
&  = & | 1 - \exp \{i \pi (R(H(\lambda)) - R(F(\lambda))) \}| \\
& \leq & \pi |R(H(\lambda)) - R(F(\lambda))| \\
& \leq & \pi L_R |H(\lambda) - F(\lambda)| \\
& \leq & \pi^2 L_R \underset{0 \leq \lambda \leq \pi} \sup |h(\lambda) - f(\lambda)|/\langle f \rangle_{\pi} \\
& = & \sqrt{\delta},
\eeqa
where we have used the fact that $|1 - e^x| \leq |x|$ for any complex number $x$.
 Because $\Psi_h$ is an all-pass
filter, by Remark \ref{rem:lip-all-pass}
 $\mbox{LIP}(\Psi, f)$ takes the form (\ref{eq:lip-all-pass}), and 
\beqa
\mbox{LIP} (\Psi_h, f)  &=& 1 - \frac{\langle \Psi_h, f\rangle^2}{ \langle f\rangle^2}\\
& = & 1 - \frac{\langle \Psi_f + \Psi_h - \Psi_f, f\rangle^2}{ \langle f\rangle^2}\\
&=& 1 - \frac{[\langle \Psi_f, f\rangle + \langle \Psi_h - \Psi_f, f\rangle ]^2}{ \langle f\rangle^2}\\
&=& 1 - \frac{ \langle \Psi_h - \Psi_f, f\rangle^2}{ \langle f\rangle^2}\\
&\geq & 1 - \frac{ \langle |\Psi_h - \Psi_f|, f\rangle^2}{ \langle f\rangle^2}\\
&\geq & 1 - (\sqrt{\delta})^2\\
& = & 1 - \delta,
\eeqa
where the fourth equality is obtained by using the fact $\langle \Psi_f , f\rangle = 0$ 
 (this is shown in the proof of Theorem 
 \ref{thm:perfect_prvcy}).
\end{proof}

\noindent{\bf Proof of Theorem~\ref{thm:delta_LIP}}
\begin{proof}
We will show that $\Psi_h$ is $\delta-$LIP by showing that $h$ belongs to the class \eqref{eq:f_nbhd}. From the definition it is clear that $\langle h \rangle_{\pi} = \langle f \rangle_{\pi}$
Also,
\beqa
|h - f| &=& |\langle f \rangle_{\pi}({\tilde{f}} + \Delta)/(1 + \pi\Delta) - f| \\
&=& (1 + \pi\Delta)^{-1}|f + \Delta\langle f \rangle_{\pi} - f - f\pi\Delta| \\
&=& (1 + \pi\Delta)^{-1}\Delta\langle f \rangle_{\pi} |\pi{\tilde{f}} - 1|. 
\eeqa
From Theorem~\ref{thm:deltaLIP}, for the privacy mechanism $\Psi_h$  associated with $h$ to be $\delta-$LIP we need
\[ 
\underset{0 \leq \lambda \leq \pi} \sup |h(\lambda) - f(\lambda)| \leq \frac{\sqrt{\delta}\langle f \rangle_{\pi}}{L_R \pi^{2}}.
\]
Thus, $\Psi_h$ is $\delta-$LIP if 
\[ 
\frac{\Delta}{1 + \pi\Delta} \, \underset{0 \leq \lambda \leq \pi} \sup| \pi {\tilde{f}}(\lambda) - 1| \leq \frac{\sqrt{\delta}}{L_R \pi^{2}}.
\]
or 
\[ 
\frac{\Delta}{1 + \pi\Delta}  \leq \frac{\sqrt{\delta}}{L_R \pi^{2}\underset{0 \leq \lambda \leq \pi} \sup| \pi {\tilde{f}}(\lambda) - 1|}.
\]
The inequality is satisfied by at least one $\Delta$ provided the RHS is strictly less than one and in that case any $\Delta$ such that $\Delta \leq B$ will satisfy the inequality. Since $\underset{0 \leq \lambda \leq \pi} \sup| \pi {\tilde{f}}(\lambda) - 1| > \frac{\sqrt{\delta}}{L_R \pi^{2}}$, we have $\Psi_h$ is $\delta-$LIP if $\Delta \leq B$. 

Using the bound $|e^x - 1| \geq \frac{1}{2}x$ for any complex number $x$, we have 
\be
|\Psi_h(e^{-i\lambda}) - \Psi_f(e^{-i\lambda})| \geq \frac{\pi}{2}|R(H(\lambda))  - R(F(\lambda))|.
\lb{eq:filter_diff}
\ee
Here $H(\lambda) = \int_0^{\lambda} {\tilde{h}}(\omega) d\omega$,
 where ${\tilde{h}} = (1 + \Delta\pi)^{-1}({\tilde{f}}(\lambda) + \Delta).$
Hence $H(\lambda) = (1 + \Delta\pi)^{-1}(F(\lambda) + \Delta\lambda) = (1 - \alpha)F(\lambda) + \alpha (\lambda/\pi)$ where $0 < \alpha = \frac{\Delta\pi}{1 + \Delta\pi}.$
Thus, $|H(\lambda) - F(\lambda)| = \alpha|F(\lambda) - \lambda/\pi|.$ 
Then from \eqref{eq:filter_diff} and the mean value theorem, we obtain the theorem's final result. 
\end{proof}

\noindent{\bf Proof of Theorem~\ref{thm:trend_delta_LIP}}
\begin{proof}
Since $\phi(z) = i\pi R(H(\lambda) )$ and $H(0) = 0$,
 we can differentiate successively, 
using the chain rule,  at $\lambda = 0$, and the 
 condition \eqref{eq:vanish_der} holds. 
 Hence the result follows.
\end{proof} 
\newpage

\bibliographystyle{plain}
\bibliography{references}

\end{document}